\documentclass[12pt]{article}
\usepackage[english]{babel}
\usepackage{ inputenc,amsmath,amsthm}
\usepackage{amsfonts}
\usepackage[margin=1in]{geometry}

\theoremstyle{definition}

\theoremstyle{remark}

\numberwithin{equation}{section}

\usepackage{amsmath}
\usepackage{graphicx}
\usepackage{comment}

\usepackage{xr}
\usepackage{setspace}

 \usepackage[symbol]{footmisc} 

\title{Improved Small Area Inference from Data Integration Using Global-Local Priors}%
\usepackage{bigfoot}

\DeclareNewFootnote{AAffil}[arabic]
\DeclareNewFootnote{ANote}[fnsymbol]

\usepackage{etoolbox}
\makeatletter
\patchcmd\maketitle{\def\@makefnmark{\rlap{\@textsuperscript{\normalfont\@thefnmark}}}}{}{}{}
\makeatother

\makeatletter
\def\thanksAAffil#1{
  \footnotemarkAAffil\protected@xdef\@thanks{\@thanks%
        \protect\footnotetextAAffil[\the \c@footnoteAAffil]{#1}}%
}
\def\thanksANote#1{%
  \footnotemarkANote%
  \protected@xdef\@thanks{\@thanks%
        \protect\footnotetextANote[\the \c@footnoteANote]{#1}}%
}
\makeatother

\author{
  Dexter Cahoy%
  \thanksAAffil{University of Houston System, Houston, Texas, 77002} 
  , %
  Joseph Sedransk%
\thanksAAffil{University of Maryland, College Park, Maryland, 20742}
}
\date{\today}

\usepackage{parskip}
\usepackage{natbib} 
\usepackage[plainpages=false,pdfpagelabels,hypertexnames,linktocpage, colorlinks, citecolor=red,linkcolor=blue]{hyperref}

\numberwithin{table}{subsection}
\numberwithin{figure}{subsection}
\begin{document}

\setlength{\parskip}{0pt} 
\setlength{\parindent}{0pt}

\maketitle

\begin{abstract} 

We present and apply methodology to improve inference for small area parameters by using data from several sources. This work extends \cite{cas23} who showed how to integrate summary statistics from several sources. Our methodology uses hierarchical global-local prior distributions to make inferences for the proportion of individuals in Florida’s counties who do not have health insurance. Results from an extensive simulation study show that this methodology will provide improved inference by using several data sources. Among the five model variants evaluated the ones using horseshoe priors for all variances have better performance than the ones using lasso priors for the local variances.

\end{abstract}

 \newpage
\section{Introduction and Background}

Significant changes in survey sampling such as greatly reduced response rates and budgets have increased attention to developing and using new methodology to improve inferences. One promising approach is to add data from other sources to the data from a single, probability-based survey.  \cite{cas23} presented two methods for combining summary statistics from several sample surveys and applied them to data from the Behavioral Risk Factor Surveillance System (BRFSS) and the Small Area Health Insurance Estimates (SAHIE)  Program  at the US Census Bureau. These methods have more structure than those commonly used in survey sampling, and \cite{cas23}   showed the value of these methods in pooling such data. They also noted the potential value of this methodology in combining data from traditional sample surveys as well as non-probability surveys and, perhaps, administrative data. 
\cite{cas23} made inference for each (Florida) county using the data only from that county, so may have lost the
opportunity for further improvements in inference by combining the data across the counties. In the current paper we extend this work by providing methodology to make simultaneous inference for all counties when there are several data sources. However, the primary methodology in \cite{cas23}, Uncertain Pooling \citep{eas01,mas92} cannot, at present, accommodate the computational demands of inference for many small areas together with several data sources. Thus, we use global-local (GL) models, a promising alternative with an extensive literature. There is a thorough summary in the review paper,
\cite{bdpw19}. There are two papers in the survey sampling literature,  
\cite{tghs18} and \cite{tg23}. Results
in \cite{tghs18} motivated our choice to use GL models.  Since both papers address a simpler case than we do, i.e., a single data source, we have extended the model in \cite{tghs18}.

A  simplified version of the Fay - Herriot model that \cite{tghs18} use is

\begin{equation} \label{FH model}
y_i = \theta_i + e_i, \quad \theta_i = \eta + u_i, \quad  i = 1,...,I
\end{equation}

where $y_i$ is the direct estimate for small area $i$ and $\eta$ is a constant. Here, $e = (e_1,..., e_I)^{'}$ and $u = (u_1,...,u_I)^{'}$ are assumed to be independent. The elements of $e$ are independent with $e_i \sim N(0, V_i)$. \cite{tghs18}  note that $V_i$ is assumed to be known to avoid problems with identifiability. The GL shrinkage prior for the $u_i$ is

\begin{equation} \label {One level shrinkage prior}
u_i | \lambda^2_i, \tau^2 \sim N(0, \lambda^2_i \tau^2).
\end {equation}

Instead of (\ref{FH model}) and (\ref{One level shrinkage prior}) \cite{tghs18} replace $\eta$ with a linear regression. They assume  a locally uniform prior on the regression parameter, and consider several priors for the variances; see their Table 1.
As seen in (\ref{One level shrinkage prior}) there are two levels for the prior variances. The global variance, $\tau^2,$ 
provides shrinkage for all of the random effects while the local variances, $\lambda_i^2,$ provide adjustments for the
individual areas. If the priors are heavy-tailed, effective inference for both small and large random effects can be made.
\cite{tg23} extend \cite{tghs18} by replacing (\ref{One level shrinkage prior}) with a normal distribution with spatial
structure. Another alternative to (\ref{One level shrinkage prior}) is a spike-and-slab prior \citep{dm15}, compared with the GL prior
in \cite{tghs18}. That is, $u_i = \xi_i v_i$ where $\xi_1,...,\xi_I$ are i.i.d. Bernoulli random variables with $P(\xi_i = 1)
= \gamma$ and $v_i \sim \text{N} (0, \zeta^2)$ independently.  A recent paper, \cite{sag23}, has several features
that are similar to our extension of \cite{tghs18} to inference on small areas when there are several sources. See Section 2 for details.

\bigskip

The rest of the paper is structured as follows. Section 2 describes our models and gives an explanation of our point estimator, the posterior mean for each FL county. Our first analysis, in Section 3,  uses the data from SAHIE and BRFSS that \cite{cas23} used to make inference about the proportion of adults without health insurance in the Florida counties. There are brief descriptions of the BRFSS survey and SAHIE program in Section 3.1.  Credible intervals for the Florida county means are in Section 3.2 while  the results of a comparison of several models using the SAHIE and BRFSS data are in Section 3.3.  Section 3.4 is about the gains from using data from two sources rather than a single source. Section 4 has the results from an extensive simulation study based, in part, on the observed data set. There is also discussion of literature
concerning the use of horseshoe and (Bayesian) lasso prior distributions. Section 5 has a discussion about the sampling variances, use of a larger number of data sources, model comparison and robustness,
concluding with a summary.

\section{Models and Inference}

Consideration of the two-source data for each of the counties in Florida suggested the following
natural extension of the model in \cite{tghs18}.   Let $Y_{ij}$ denote the value of $Y$ for the $j$-th source in the $i$-th county.  For a fixed $V_{ij},$ 

\begin{align} 
& Y_{ij}| \theta_{ij} , V_{ij} \stackrel{ind}{ \sim} N(\theta_{ij}, V_{ij}): j = 1,...,J; i = 1,...,I, \nonumber\\
&\; \theta_{ij}| \mu_i, \lambda^2_{ij},\lambda^2_i, \tau_1^2 \stackrel{ind}{\sim}N( \mu_i,  \lambda^2_{ij} \lambda^2_i \tau_1^2 ), \nonumber\\
&\mu_i | \eta,  \lambda^2_i, \tau_2^2 \stackrel{ind}{\sim} N(\eta,  \lambda^2_i \tau_2^2 ),\nonumber\\
&f(\eta) = \text{constant}. \label{M11}
\end{align}

As in \cite{cas23}, $Y$ is an estimate of the population proportion without health insurance. Note that the data 
available from BRFSS and SAHIE are only $Y$ and the estimated standard error of $Y.$ Here, inference for the county means, $\mu_i,$ is of primary interest.

Defining $\lambda = (\{\lambda^2_{ij}: j = 1, ..., J, i = 1,...,I; \lambda^2_i : i = 1,...,I\})^{'}$ and $\tau = (\tau^2_1, \tau^2_2)^{'}$, the joint prior density is
$$f(\lambda, \tau) \propto [\prod_{i,j} f(\lambda^2_{ij})][\prod_i f(\lambda^2_i)] f(\tau^2_1)f(\tau^2_2).$$

To choose the priors for the local and two global variances we have relied on discussion in the literature.
\cite{tghs18} note that \cite{ge06} and \cite{ps12} prefer horseshoe priors for top level scale parameters, and there
seems to be substantial additional support for this choice: for more details see Section 4.3.

A recent paper by \cite{sag23} uses a somewhat similar model in
a different context, estimating demand for large assortments of differentiated
goods. Their model is structured around a graphical representation of a product
classification tree. In this tree, the lowest level represents the most detailed
definition of product groups, while the highest level corresponds to the broadest
categories. Using this tree as a framework, the authors develop hierarchical
models that directly shrink product-level elasticities towards higher-level price elasticities. The structure in our problem is quite different, i.e., nesting of sources
$(j)$ within counties $(i)$. Another assumption indicates that the objectives of the
two papers are different. In the context of small area inference taking $\eta$ to have a locally uniform distribution is the natural choice while \cite{sag23} assume a standard normal distribution at the highest level. Another difference is that \cite{sag23} only consider hierarchical structure for the local
variances, i.e., $\lambda^2_{ij}\lambda^2_i$
in (\ref{M11}), a limitation since we have found better performance
for our problem using the non-hierarchical $\lambda^2_{ij}$ in the conditional variance of
$\theta_{ij}$. However, we have used ideas in \cite{sag23} in designing our
simulation study, and clarifying some features of our approach.

\bigskip

Let $\upsilon$ denote a generic variance and
$\kappa \sim \mathcal{C}^+ (0,1)$, the
half-Cauchy distribution with pdf $f(y) = 2/\pi (1+ y^2), y \ge 0.$ Then $\upsilon = \kappa^2$ has the \emph{horseshoe} distribution with
pdf $f(\upsilon) \propto [(1 + \upsilon^2) \sqrt{\upsilon^2}]^{-1}$ for $\upsilon > 0.$ The \emph{lasso} distribution has
pdf $f(\upsilon) \propto \text{exp}(-\upsilon)$ for $\upsilon \ge 0.$

Our choices are
\begin{align*}
& \text{M11a:} \quad \lambda^2_{ij}, \lambda^2_i, \tau^2_1, \tau^2_2 \quad \text{have horseshoe distributions}. \\
& \text{M11b is the same as M11a, except that} \quad \lambda^2_{ij} \quad \text{and}\quad \lambda^2_i  \quad \text{have lasso distributions.}\\
& \text{M1a is the same as M11a, except that} \quad \theta_{ij} \sim \text{N}(\mu_i, \lambda^2_{ij}\tau^2_1).\\
& \text{M1b is the same as M1a, except that} \quad \lambda^2_{ij} \quad \text{and} \quad  \lambda_i^2 \quad \text{have lasso distributions.} \\
& \text{M12 has} \quad \lambda^2_{ij} = \lambda^2_i = 1. \text{This is the usual hierarchical model.}
\end{align*}

The first models that we chose were M1a and M1b with $\theta_{ij} \sim \text{N} (\mu_i, 
 \lambda^2_{ij}\tau^2_1).$
We added the more general M11a and M11b (with variance $\lambda^2_{ij}\lambda^2_i\tau^2_1),$ after reading \cite{sag23}. 

\
\bigskip

\bigskip

To investigate the posterior mean of $\mu_i$ given the observed data and variances assume M11a and define
$A_i = \lambda^2_i \tau^2_2,  \xi_{ij}^2 = V_{ij} + \lambda^2_{ij}\lambda^2_i\tau_1^2, 
\eta_i^2 = (\sum_{j = 1}^J \xi_{ij}^{-2})^{-1}$ and $\phi_i = A_i/(A_i + \eta_i^2).$ 

 Defining
$\bar{y}_i = \sum_{j = 1}^J (y_{ij}/\xi_{ij}^2)/\sum_{j = 1}^J (1/\xi_{ij}^2)$,
the posterior mean of $\mu_i$  is

\begin{equation} \label{posterior mean}
E(\mu_i | y, \Omega) = \phi_i \bar{y}_i + (1 - \phi_i) \bar{y}_w
\end{equation}
where $\Omega = (\lambda, \tau, \{V_{ij} : j = 1,..., J; i = 1,...,I\})$, $y = \{y_{ij}: j = 1,...,J; i = 1,...,I\}$
and $\bar{y}_w =  \sum_{i = 1}^I (\bar{y}_i/(A_i + \eta_i^2))/\sum_{i = 1}^I (1/(A_i + \eta_i^2)). $

\bigskip

The \emph{within} county weight, $\xi_{ij}^{-2}/\sum_{j =1}^J \xi_{ij}^{-2},$ reduces to $V_{ij}^{-2}/\sum_{j =1}^J V_{ij}^{-2},$  if $\lambda_{ij}^2 \lambda_i^2\tau^2_1 = 0.$
If not, the local and global variance components will accommodate observed values that do not conform to the
standard hierarchical model M12. (Hereafter, we refer to such observations as ``aberrant.") Increasing $V_{ij}$, being an aberrant source (large $\lambda_{ij}^2$), or an aberrant county (large $\lambda_i^2)$ will reduce the weight on $y_{ij}$. If $V_{ij}$ is small relative to $\lambda_{ij}^2\lambda_i^2\tau_1^2, \quad \xi_{ij}^2 = \lambda_{ij}^2\lambda_i^2\tau_1^2$ which
implies that $\bar{y}_i =\sum_{j=1}^J (y_{ij}/\lambda_{ij}^2)/\sum_{j=1}^J (1/\lambda_{ij}^2).$
$A_i$ is a measure of variability \emph{across} counties while $\eta_i^2$ is a measure of variability \emph{within} counties (sum of sampling variance and variance across sources within counties). So, $\phi_i = A_i/ (A_i + \eta_i^2)$ is the standard overall shrinkage factor. Finally, $\bar{y}_w$
is a weighted average of the $\bar{y}_i$ where the weights include the adjustments based on the local variances $\lambda_i^2$ and $\lambda^2_{ij}.$

\section{Analysis of Data from Florida Counties}

\subsection{Background}
One data source is the Small Area Health Insurance Estimates Program.  (The data can be obtained from \url{https://www.census.gov/data/datasets/time-series/demo/sahie/estimates-acs.html.}) The SAHIE program uses point estimates from the American Community
Survey (ACS) together with administrative data such as Federal income tax returns and Medicaid/Children's Health Insurance Program (CHIP) participation rates. There is detailed area level modelling. The second source is the Behavioral Risk
Factor Surveillance System, obtained through telephone interviews. (The data can be downloaded from \url{https://www.flhealthcharts.gov/.}) It uses a disproportionate stratified
sample design. For additional details see Sections 3 and 7 of \cite{cas23}. For consistency with \cite{cas23} we use
2010 data for each source. The BRFSS and SAHIE data available for each Florida county are an estimate of the population proportion
uninsured ($Y$) and the estimated standard error of $Y$.

Since the analysis of the Florida county data is straightforward we present it first. As such it provides the background for the detailed simulation study in Section 4.

\subsection{Inference}

\begin{figure}[h!t!b!p!]
\centering
\includegraphics[height=8.75in, width=6.65in]{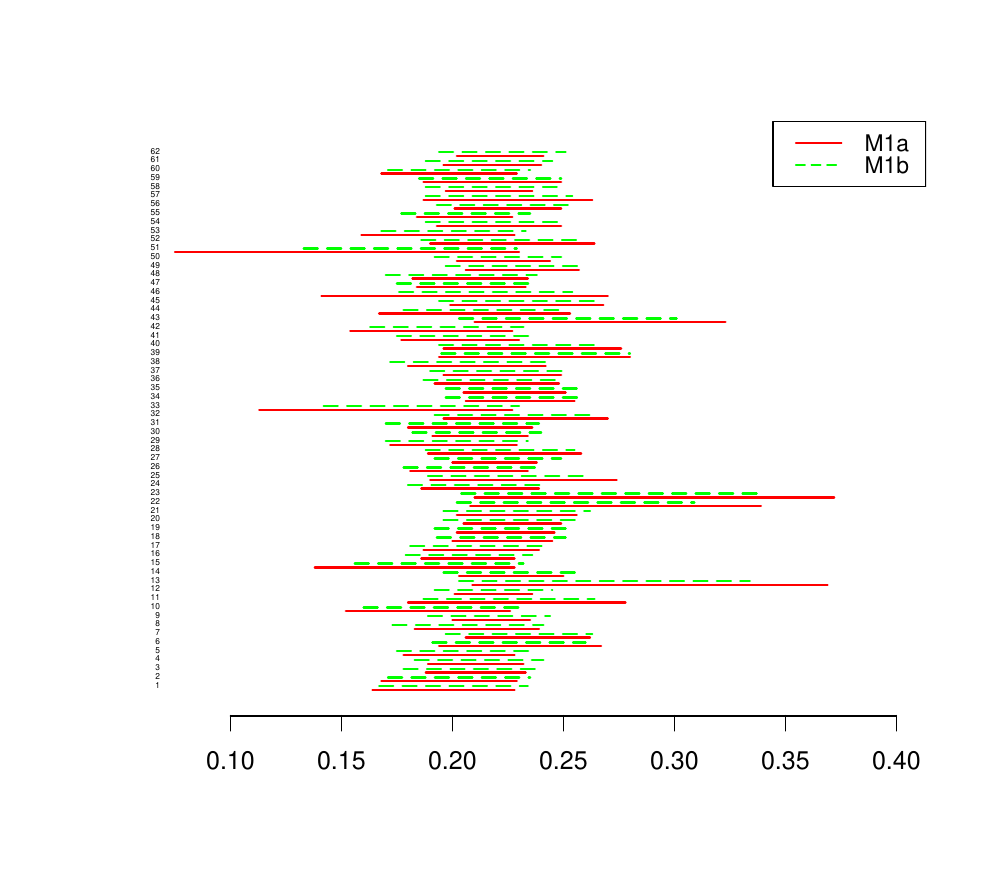} 
\caption{\label{fig:M195CI}  The 95\% credible intervals for the 62 county means using M1a and M1b.}
 \end{figure}

Figure~\ref{fig:M195CI} gives the 95\% credible intervals for the Florida counties corresponding to models M1a and M1b. Overall, the pairs of intervals have similar centers:   the two distributions, taken over the $I$ counties, are almost the same (first quartiles, medians and third quartiles are within 0.002). While the median interval length is smaller for M1a (0.056) than M1b (0.064) there are eight counties where the intervals from M1a are substantially \emph{wider}. These are situations where both the BRFSS and SAHIE point estimates are outliers or where there is a substantial difference between the two estimates. This is a first indication that M1a (with horseshoe priors for all variances) provides more appropriate inferences when there are aberrant observations. Comparing models M11a and M11b the conclusions are similar.

\subsection{Model comparisons}

\begin{figure}[h!t!b!p!]
\hspace{-0.5in}
\centering
\includegraphics[height=3.75in, width=7in]{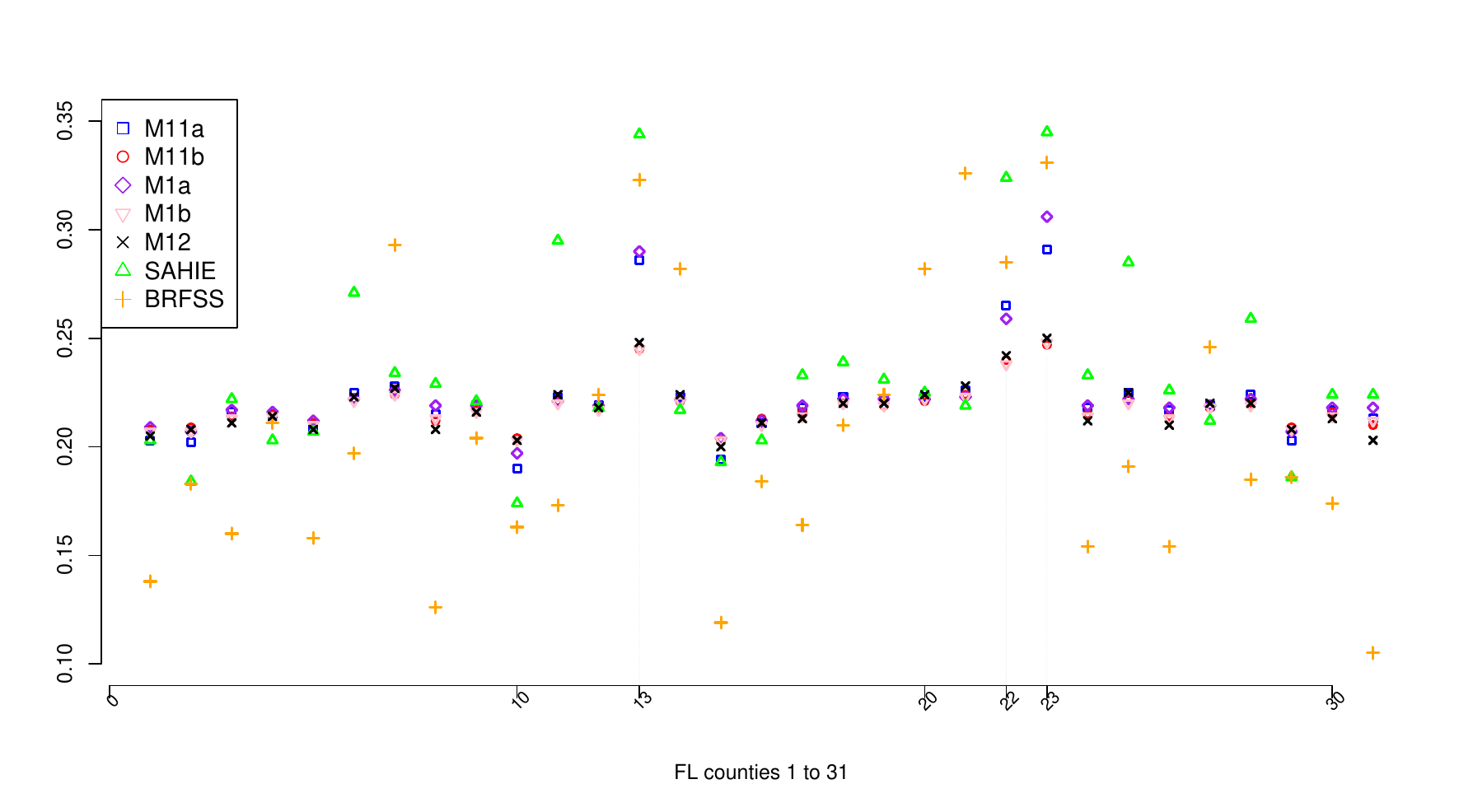} 
\caption{\label{fig:alla}  Posterior means, BRFSS and SAHIE estimates for FL counties 1-31.}  
 \end{figure}

\begin{figure}[h!t!b!p!]
\hspace{-0.5in}
\centering
\includegraphics[height=3.75in, width=7in]{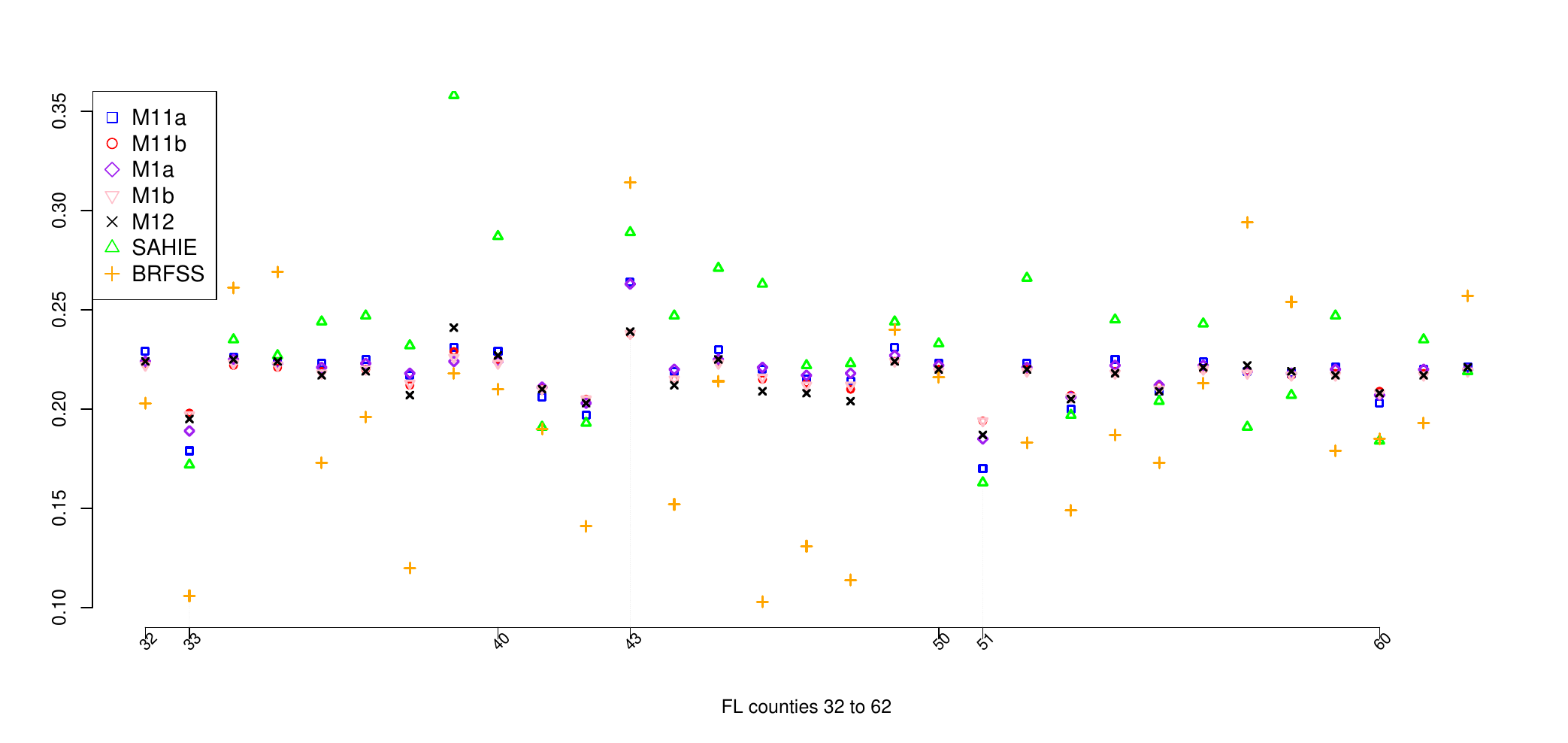} 
\caption{\label{fig:allb}  Posterior means, BRFSS and SAHIE estimates for FL counties 32-62.}  
\end{figure}

Figures  \ref{fig:alla}  and \ref{fig:allb} give the posterior mean, $\mu_i$, for each Florida county for models M11a, M11b, M1a, M1b, the standard model M12, and the two point estimates.  
First, note that there is little variation in M12 across the counties. For most counties there is agreement among the five models.   M1b and M11b (both lasso) have values of the posterior mean much closer to M12 than M1a and M11a (both horseshoe). This is most evident for the counties with both high and low values of the BRFSS and SAHIE point estimates such as 10, 13, 22 and 23 (see Figure \ref{fig:alla}).  The values for M11b and M1b are similar while, typically, the values for M11a are a little more extreme than those for M1a. 

For most counties there is not much difference between M1a and M1b. Generally, M1b is closer than M1a to the standard model M12. There are substantially more counties where M12 is closer to SAHIE than BRFSS. 
When one of SAHIE and BRFSS is small and the other is large, M1a and M1b tend to be similar and within the range defined by SAHIE and BRFSS. The most important finding is that when both SAHIE and BRFSS are large or small M1b shrinks much more than M1a. Similarly, M11b shrinks more than M11a. This is in accord with the literature. For their one stage model, i.e., without source effects, \cite{tghs18} say that the``shrinkage factors under[lasso] priors are stochastically larger than those under [horseshoe] priors thus causing more shrinkage.'' \cite{sag23} add that ``the tails of an exponential mixing density are lighter than the polynomial tails of the half-Cauchy, suggesting that the Bayesian lasso may tend to over-shrink large regression coefficients and under-shrink small ones relative to the horseshoe.''




As seen in (\ref{posterior mean}) the magnitude of $\phi$ determines the contribution of the observations from county $i$ to the posterior mean of $\mu_i.$ Figures \ref{fig:M1sphi} - \ref{fig:M12phi} in the appendix show the posterior distributions of $\{\phi_i : i = 1,...,I\}$ for models M1a, M1b, M11a, M11b and M12. These figures display considerable differences among the models. For M12 the set of $I$ distributions of $\phi$ show little variation which is markedly different from those for the four GL models. Comparing M1a (with horseshoe priors for all variances) with M1b (with lasso priors for the local variances and horseshoe priors for the global variances) there are notable differences, i.e., much greater variation for M1a both across and within counties. This is also seen for the counterparts of M1a and M1b, M11a and M11b, which differ from M1a and M1b by having local variances $\lambda^2_{ij}\lambda^2_i$ rather than $\lambda^2_{ij}.$

\subsection{Two sources vs. one source}

We use the Florida county data set to illustrate differences in inferences when there is only one data source rather than the two data sources. The one source model is

$$Y_i | \mu_i, V_i \stackrel {ind}{\sim} N(\mu_i, V_i),$$
$$\mu_i | \eta \stackrel {ind}{\sim} N(\eta, \lambda_i^2 \tau^2),$$
$$f(\eta) =\text{ constant}.$$

\begin{figure}[h!t!b!p!]
\hspace{-0.25in}
\centering
\includegraphics[height=3.75in, width=6.85in]{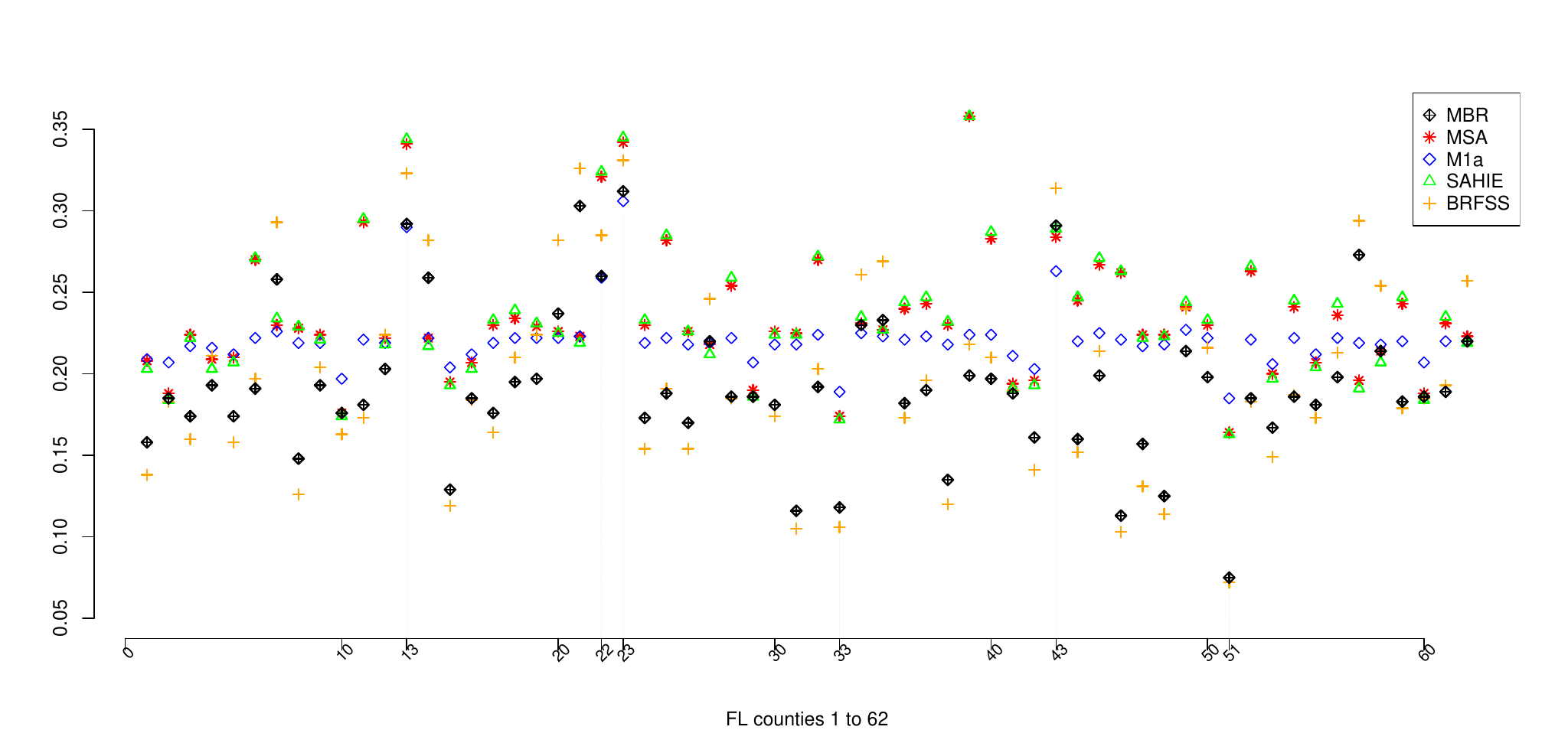} 
\caption{\label{fig:M1a}   MBR, MSA,  M1a,   observed BRFSS and SAHIE estimates.} 
 \end{figure}

The most meaningful comparison is with the BRFSS data set since it is the one with much larger observed standard errors, thus potentially benefiting from the addition of the SAHIE data when this is appropriate. One may consider
the analysis in this section 
as an illustration or as a realistic choice because of (apparently) untested assumptions made in SAHIE.  For each county
Figure \ref{fig:M1a} gives the BRFSS and SAHIE point estimates and posterior means corresponding to M1a (horseshoe) and the one source horseshoe models (MBR, MSA) applied to the BRFSS and SAHIE data.

Typically, the BRFSS posterior mean (using the one source horseshoe model) adjusts the BRFSS point estimate by increasing smaller values and decreasing larger values of the latter. The M1a posterior mean adjusts the one source BRFSS posterior mean by increasing smaller values and
decreasing larger values of the one source BRFSS posterior mean. With this data set the M1a posterior mean doesn't vary very much over the counties, so there are adjustments for a fairly small set of counties. But note that M1a has excellent
properties, as shown in the results from the simulation study (Section 4).

\begin{figure}[h!t!b!p!]
\hspace{-0.25in}
\centering
\includegraphics[height=3.75in, width=6.85in]{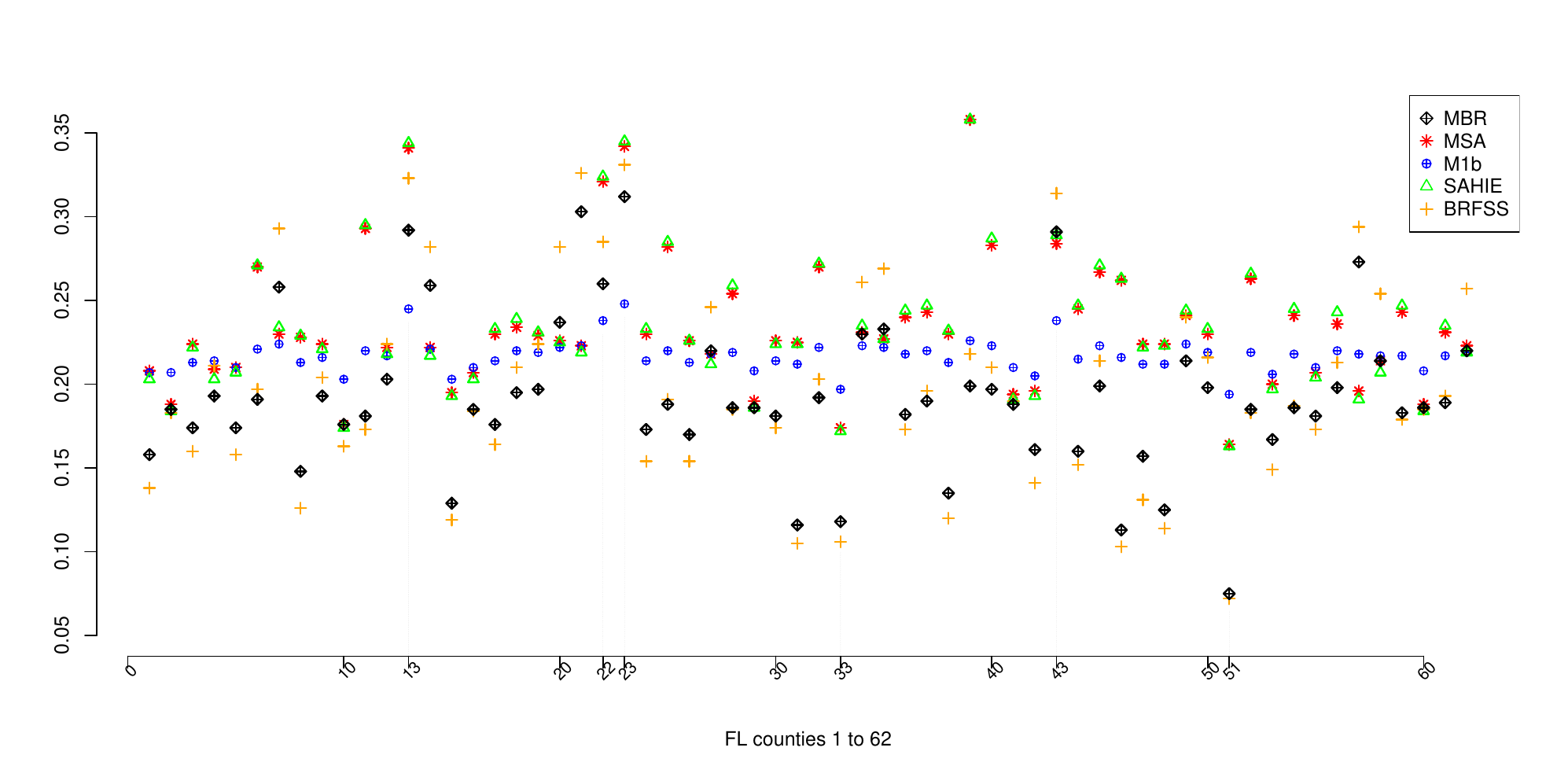} 
\caption{\label{fig:M1b}  MBR, MSA,  M1b,   observed BRFSS and SAHIE estimates.}  
 \end{figure}

The results for M11a are similar to those for M1a.  This is not surprising since M11a and M1a differ only in the variance of $\theta_{ij}$ (see Section 2).  With one important exception the conclusions above hold for M11b and M1b, both of which use lasso priors. Unlike for M11a and M1a,  there is considerable shrinkage towards the common county value even for some counties (e.g., 13, 22, 23, 33, 43, 51)  with somewhat extreme values of the BRFSS point estimate: see  Figure \ref{fig:M1b} which has the same format as   Figure \ref{fig:M1a}.  

\bigskip

\section{Simulation Study}

\subsection{Introduction}

Our simulation study is patterned after those in \cite{tghs18} and \cite{sag23}. In this section we compare the performance of models M1a, M1b, M11a and M11b with the base method M12. (See Section 2 for the definitions of these models.)  We present the results obtained by generating data from the following models. 

\bigskip

Let $Y_{ij}$ denote the value of $Y$ for the $j$-th source in the $i$-th county. For a fixed $V_{ij}$ 

\[ \label{Y equation}
Y_{ij} \stackrel{ind}{\sim} N(\theta_{ij}, V_{ij}) \; :  j = 1,...,J; i = 1,...,I,
\]
\[\label{theta equation}
\theta_{ij} \stackrel {iid}{\sim} N(\mu_i, \gamma^2_1),
\]
\begin{equation} \label{mu equation}
\mu_i \stackrel{iid}{\sim} N(\eta, \gamma^2_2). 
\end{equation}

For the $\gamma^2,$ we have used mixture and outlier representations. For a mixture,
$\gamma^2 = \delta_1 \tau^2_{21} + (1 - \delta_1)\tau^2_{22}$,
$\delta_1 \in \{0,1\}$ with $P(\delta_1 = 1) = p_1$ and $\tau^2_{22} = 0.05^2$.
For an outlier $\gamma^2 = \delta_2 \tau^2_{11} $ with $\delta_2 \in \{0,1\}$ and $P(\delta_2 = 1) = p_2.$
With a large number of quantities to specify we have simplified by not indexing by county or
source. There are exceptions, i.e., for cases 5 and 6, described in Section 4.5.

To anchor this simulation study, we have chosen values of some quantities based
on the BRFSS and SAHIE data: $\eta$ is a constant, $0.25,$ with $I = 62$ Florida counties and $J = 2$ sources.
Also, the sampling variances, $V_{ij}$,  are the values observed in these surveys.  Section 5 has further discussion about sampling variances, including the issue of identifiability noted by \cite{tghs18}.

\bigskip

Case 1 has both $\gamma$'s as outliers while Case 3 has both as mixtures. For Case 2 $\gamma^2_1$ is an outlier
while $\gamma^2_2$ is a mixture. For Case 4 $\gamma^2_1$ is a mixture
while $\gamma^2_2$ is an outlier. We considered many values of the probabilities ($p_1, p_2$) and variances
($\tau^2$) and present those that are the most informative.

For each of the four cases and choice of the probabilities and the variances we generated 100 data sets. For each data set we estimated $\mu_i$ by its posterior mean, $\hat{\mu}_i$,
and evaluated the fit using four common measures, i.e., average absolute relative deviation (ARB), average squared
relative deviation (ASRB), average absolute deviation (AAD) and average squared deviation (ASD). We summarize for each data set as $\text{ARB} = I^{-1}\sum_{i=1}^I\frac{(\lvert\hat{\mu}_i - \mu_i\lvert)}{\mu_i},
\text{ASRB} =  I^{-1}\sum_{i=1}^I\frac{(\hat{\mu}_i - \mu_i)^2}{\mu_i^2},
\text{AAD} =  I^{-1}\sum_{i=1}^I\lvert\hat{\mu}_i - \mu_i\lvert, \quad \text{and} \quad
\text{ASD} =  I^{-1}\sum_{i=1}^I(\hat{\mu}_i - \mu_i)^2$ where $I$ is the number of Florida counties. For each measure (ARB, ASRB, AAD, ASD) the final summary statistic is the median from the 100 data sets. To simplify, we present results only for ARB and ASRB.

\subsection{Posterior computation}

All of the calculations were implemented in R using the  Advanced Cyberinfrastructure Coordination Ecosystem: Services \& Support (ACCESS)'s Pittsburgh Supercomputing Center.    We generated 100 data samples and calculated posterior estimates using 18K MCMC runs and 3K burn-ins for each sample. To test convergence, we simulated five chains of length 7K (with 2K burn-ins) for the posterior distributions of the $\mu_i$'s, and obtained values of the split-R variant of Gelman's diagnostic close to 1.0 for all 62 counties.  Section 11.4 of \cite{getal13} has a discussion of convergence measures including the split-R
variant.  The full conditional distributions  for models M11a and M11b are given in Appendix \ref{cond}. The corresponding Gibbs algorithms for M1a and M1b can be obtained by simplifying the given expressions.

\subsection{Overall evaluation}

Our comparisons are made relative to the standard model, M12, as discrepancy ratios. Model 12 is given in Section 2: it is the special case of the global-local model with $\lambda_{ij}^2 = \lambda_i^2 = 1.$ For model Mx (x=1a, 1b, 11a, 11b) and ARB the discrepancy ratio is denoted by ARB(Mx)/ARB(M12) with small values of the ratio showing gains for Mx relative to M12.  We summarize the distributions (over the set of specifications) of the discrepancy ratio for models M1a, M1b, M11a and M11b using the medians.

For Cases 1 - 4 the medians of ARB(M1a)/ARB(M12), ARB(M11a)/ARB(M12), \linebreak ARB(M1b)/ARB(M12)
and ARB(M11b)/ARB(M12) are (0.589, 0.633, 0.802, 0.853), (0.733, 0.760, 0.851, 0.874), (0.780, 0.821, 0.870, 0.915), and (0.786, 0.913, 0.835, 0.929). They show considerable gains for all of the global-local models with M1a having the smallest values (largest gain). The results for ASRB are similar.

Additional information is in Tables  \ref{table:summ1} -\ref{table:summ4} 
in the Supplementary Material  corresponding to the four cases. In each table we present for ARB and ASRB,
and each of the four discrepancy ratios, the minimum value, first quartile, median, mean, third quartile and maximum value 
(taken over the set of specifications).

Table \ref{table:counts1}  gives an overall summary. For each discrepancy ratio, case and model, Mx, the cell entry is the number of specifications where Mx has the smallest value among the set of models. 
Clearly, model M1a is preferred. Moreover, M1a has a smaller value of ARB(M1a)/ARB(M12) than ARB(M1b)/ARB(M12) for 99
of the 122 settings. The other horseshoe model, M11a, has a smaller value of ARB(M11a)/ARB(M12) than ARB(M11b)/ARB(M12)
for 95 settings. The results for ASRB show a greater preference for these horseshoe models. See Table \ref{table:counts2}  for a breakdown by case.  

\bigskip

These results lend additional credence to the literature showing a preference for the horseshoe prior.
For example,  \cite{sag23} say that ``the tails of an exponential mixing density are lighter than the polynomial tails of the half-Cauchy, suggesting that the Bayesian lasso may tend to over-shrink large regression coefficients and under-shrink small ones relative to the horseshoe.'' \cite{mas16} add that the ``choice of a half-Cauchy prior distribution over the global and local hyperparameters results in aggressive shrinkage of small coefficients (i.e., noise) and virtually no shrinkage of sufficiently large coefficients (i.e., signal). This is contrast to the well-known Bayesian lasso ... and Bayesian ridge hierarchies where the shrinkage effect is uniform across all coefficients.'' 

For the simpler model in (\ref{FH model}) and (\ref{One level shrinkage prior}) \cite{bdpw19} show why the horseshoe prior is effective. This
argument also holds for inference about the source effects in our (\ref{M11}), since the first two expressions in (\ref{M11})
are essentially the same as the model in (\ref{FH model}) and (\ref{One level shrinkage prior})  (drop the subscript $i$). For our preferred
model, M1a, take $\lambda^2_i = 1.$ Then

\begin{equation}
E(\theta_{ij}|y_{ij}) = [1 - E(\kappa_{ij}| y_{ij})]y_{ij} + E(\kappa_{ij}|y_{ij})\mu_i
\end{equation}
where the shrinkage weight, $\kappa$, is $\kappa_{ij} = V_{ij}/(V_{ij} + \lambda^2_{ij}\tau^2_1).$
\cite{bdpw19} write the marginal likelihood of $y_{ij}$ in terms of $\kappa_{ij},$ noting that the posterior
density of the shrinkage weight identifies signals and noise by letting $\kappa_{ij} \rightarrow 0$ and $\kappa_{ij} \rightarrow 1.$ The marginal likelihood is zero when the shrinkage weight is zero, so it does not help identify the signals.
However, as shown in \cite{bdpw19} using the horseshoe prior enables the posterior to approach both zero and one.
This gives an indication why M1a, with horseshoe prior at the source level, has better performance than M1b, with
lasso prior at the source level.

\bigskip

\begin{table}[htbp]
\centering
\begin{tabular}{ccccc}
   &   & ARB &   &  \\ 
  \hline
  Case & M11a & M11b & M1a & M1b \\ 
   1 & 0 & 1 & 29 & 0 \\ 
   2 & 8 & 0 & 28 & 0 \\ 
  3 & 0 & 1 & 16 & 3 \\ 
  4 & 0 & 0 & 17 & 19 \\ 
   &   & &   &  \\ 
   &  & ASRB &  &  \\ 
\hline
  Case & M11a & M11b & M1a & M1b \\ 
  1 & 5 & 0 & 15 & 0 \\ 
  2 & 6 & 0 & 30 & 0 \\ 
    3 & 9 & 0 & 8 & 3 \\ 
  4 & 2 & 1 & 21 & 12 \\ 
   \hline
\end{tabular}
\caption{\label{table:counts1} For each case: number of specifications where a model has the best performance.}
\end{table}

Tables \ref{table:results1} - \ref{table:results4}  give more detailed information for M1a. Each table refers to a case, e.g., in Table  \ref{table:results1} both $\gamma^2$'s are outliers. Each row corresponds to a specification of the probabilities and the variances, $\tau^2.$ The first set of columns identify the probabilities and the $\tau^2$, e.g., for Case 1 the components of $\gamma^2_1$ in (\ref{theta equation}) corresponding to the outlier model and the components of $\gamma^2_2$ in (\ref{mu equation}) corresponding to the outlier model.
The last two columns display the discrepancy ratios for ARB and ASRB corresponding to model M1a, i.e., ARB(M1a)/ARB(M12) and ASRB(M1a)/ASRB(M12).  Tables ~\ref{table:specs1} - \ref{table:asrb4}  
in the Supplementary Material give the same information for the other models. 
These tables can be used to investigate the effects of changing the values of the probabilities and variances: see Section 4.4.

\subsection{Changing probabilities and variances}

We next investigate the effects of changing the probabilities and values of the variances ($\tau^2$). This is a new evaluation since these quantities are fixed in \cite{tghs18} and \cite{sag23}. Motivated by the results in Section 3.3 and the literature referenced in Section 2 we only summarize our findings about properties of M1a. Values for the probabilities and 
variances were chosen, initially, based on the results in Section 3. Other values were added to try to sharpen the
conclusions. In Tables \ref{table:results1} - \ref{table:results4} we present the results for a subset of the choices of the probabilities and variances that we
considered. For the most part the gains for M1a increase as the probabilities and variances increase. There are
some exceptions, not explained by further investigation.


\begin{table}[h!t!b!p!]
\centering
\begin{tabular}{ccccccc}
 & $p_2$   for $\mu_i$ & $p_2$  for $\theta_{ij}$  & $\tau_{11}$ for $\mu_i$ & $\tau_{11}$ for $\theta_{ij}$  & M1a/M12(ARB)  & M1a/M12(ASRB) \\ 
  \hline
1 & 0.100 & 0.100 & 0.025 & 0.025 & 0.803 & 0.670 \\ 
  2 &  &  & 0.050 & 0.050 & 0.563 & 0.466 \\ 
  3 &  &  & 0.100 & 0.100 & 0.345 & 0.226 \\ 
  4 &  &  & 0.200 & 0.200 & 0.244 & 0.058 \\ 
  5 &  &  & 0.050 & 0.100 & 0.487 & 0.327 \\ 
  6 &  &  & 0.100 & 0.050 & 0.425 & 0.282 \\ 
  7 & 0.200 & 0.200 & 0.025 & 0.025 & 0.828 & 0.832 \\ 
  8 &  &  & 0.050 & 0.050 & 0.635 & 0.601 \\ 
  9 &  &  & 0.100 & 0.100 & 0.429 & 0.293 \\ 
  10 &  &  & 0.200 & 0.200 & 0.298 & 0.123 \\ 
  11 &  &  & 0.050 & 0.100 & 0.636 & 0.544 \\ 
  12 &  &  & 0.100 & 0.050 & 0.562 & 0.486 \\ 
  13 & 0.400 & 0.400 & 0.025 & 0.025 & 0.896 & 0.874 \\ 
  14 &  &  & 0.050 & 0.050 & 0.772 & 0.784 \\ 
  15 &  &  & 0.100 & 0.100 & 0.636 & 0.500 \\ 
  16 &  &  & 0.200 & 0.200 & 0.612 & 0.481 \\ 
  17 &  &  & 0.050 & 0.100 & 0.780 & 0.727 \\ 
  18 &  &  & 0.100 & 0.050 & 0.728 & 0.711 \\ 
  19 & 0.100 & 0.200 & 0.025 & 0.025 & 0.923 & 0.888 \\ 
  20 &  &  & 0.050 & 0.050 & 0.595 & 0.442 \\ 
  21 &  &  & 0.100 & 0.100 & 0.405 & 0.328 \\ 
  22 &  &  & 0.200 & 0.200 & 0.255 & 0.096 \\ 
  23 &  &  & 0.050 & 0.100 & 0.609 & 0.579 \\ 
  24 &  &  & 0.100 & 0.050 & 0.456 & 0.370 \\ 
  25 & 0.200 & 0.100 & 0.025 & 0.025 & 0.796 & 0.773 \\ 
  26 &  &  & 0.050 & 0.050 & 0.600 & 0.504 \\ 
  27 &  &  & 0.100 & 0.100 & 0.415 & 0.240 \\ 
  28 &  &  & 0.200 & 0.200 & 0.309 & 0.093 \\ 
  29 &  &  & 0.050 & 0.100 & 0.544 & 0.369 \\ 
  30 &  &  & 0.100 & 0.050 & 0.583 & 0.433 \\ 
   \hline
\end{tabular}
\caption{\label{table:results1} Simulation specifications  and ARB(M1a)/ARB(M12), ASRB(M1a)/ASRB(M12) for Case 1.}
\end{table}


\begin{table}[h!t!b!p!]
\centering
\begin{tabular}{ccccccc}
 & $p_1$ for  $\mu_i$  & $p_2$ for  $\theta_{ij}$ & $\tau_{21}$ for  $\mu_i$   & $\tau_{11}$ for  $\theta_{ij}$  & M1a/M12(ARB)  & M1a/M12(ASRB)\\ 
 
  \hline
1 & 0.100 & 0.100 & 0.100 & 0.050 & 0.942 & 0.892 \\ 
  2 &  &  &  & 0.100 & 0.725 & 0.520 \\ 
  3 &  &  &  & 0.200 & 0.502 & 0.255 \\ 
  4 &  &  & 0.200 & 0.050 & 0.894 & 0.816 \\ 
  5 &  &  &  & 0.100 & 0.723 & 0.707 \\ 
  6 &  &  &  & 0.200 & 0.551 & 0.375 \\ 
  7 &  & 0.200 & 0.100 & 0.050 & 0.936 & 0.887 \\ 
  8 &  &  &  & 0.100 & 0.709 & 0.473 \\ 
  9 &  &  &  & 0.200 & 0.487 & 0.166 \\ 
  10 &  &  & 0.200 & 0.050 & 0.887 & 0.823 \\ 
  11 &  &  &  & 0.100 & 0.739 & 0.569 \\ 
  12 &  &  &  & 0.200 & 0.498 & 0.217 \\ 
  13 &  & 0.400 & 0.100 & 0.050 & 0.951 & 0.873 \\ 
  14 &  &  &  & 0.100 & 0.742 & 0.481 \\ 
  15 &  &  &  & 0.200 & 0.525 & 0.259 \\ 
  16 &  &  & 0.200 & 0.050 & 0.923 & 0.862 \\ 
  17 &  &  &  & 0.100 & 0.754 & 0.617 \\ 
  18 &  &  &  & 0.200 & 0.556 & 0.330 \\ 
  19 & 0.200 & 0.100 & 0.100 & 0.050 & 0.959 & 0.940 \\ 
  20 &  &  &  & 0.100 & 0.713 & 0.579 \\ 
  21 &  &  &  & 0.200 & 0.483 & 0.252 \\ 
  22 &  &  & 0.200 & 0.050 & 0.924 & 0.826 \\ 
  23 &  &  &  & 0.100 & 0.702 & 0.570 \\ 
  24 &  &  &  & 0.200 & 0.519 & 0.405 \\ 
  25 &  & 0.200 & 0.100 & 0.050 & 0.943 & 0.892 \\ 
  26 &  &  &  & 0.100 & 0.764 & 0.616 \\ 
  27 &  &  &  & 0.200 & 0.465 & 0.203 \\ 
  28 &  &  & 0.200 & 0.050 & 0.886 & 0.723 \\ 
  29 &  &  &  & 0.100 & 0.749 & 0.515 \\ 
  30 &  &  &  & 0.200 & 0.492 & 0.245 \\ 
  31 &  & 0.400 & 0.100 & 0.050 & 0.917 & 0.725 \\ 
  32 &  &  &  & 0.100 & 0.775 & 0.603 \\ 
  33 &  &  &  & 0.200 & 0.521 & 0.260 \\ 
  34 &  &  & 0.200 & 0.050 & 0.919 & 0.793 \\ 
  35 &  &  &  & 0.100 & 0.727 & 0.668 \\ 
  36 &  &  &  & 0.200 & 0.572 & 0.411 \\ 
   \hline
\end{tabular}
\caption{\label{table:results2} Simulation specifications  and ARB(M1a)/ARB(M12), ASRB(M1a)/ASRB(M12) for Case 2.}
\end{table}


\begin{table}[h!t!b!p!]
\centering
\begin{tabular}{ccccccc}
& $p_1$  for $\mu_i$ & $p_1$ for $\theta_{ij}$ & $\tau_{21}$ for $\mu_i$ & $\tau_{21}$  for $\theta_{ij}$ & M1a/M12(ARB)  & M1a/M12(ASRB)\\ 
  \hline
1 & 0.100 & 0.100 & 0.100 & 0.100 & 1.050 & 1.102 \\ 
  2 &  &  & 0.200 & 0.200 & 0.844 & 0.452 \\ 
  3 &  &  & 0.400 & 0.400 & 0.655 & 0.305 \\ 
  4 &  &  & 0.200 & 0.400 & 0.680 & 0.214 \\ 
  5 &  &  & 0.400 & 0.200 & 0.828 & 0.562 \\ 
  6 & 0.200 & 0.200 & 0.100 & 0.100 & 1.001 & 0.960 \\ 
  7 &  &  & 0.200 & 0.200 & 0.766 & 0.470 \\ 
  8 &  &  & 0.400 & 0.400 & 0.636 & 0.427 \\ 
  9 &  &  & 0.200 & 0.400 & 0.619 & 0.275 \\ 
  10 &  &  & 0.400 & 0.200 & 0.783 & 0.763 \\ 
11 & 0.100 & 0.200 & 0.100 & 0.100 & 1.025 & 0.997 \\ 
  12 &  &  & 0.200 & 0.200 & 0.786 & 0.527 \\ 
  13 &  &  & 0.400 & 0.400 & 0.619 & 0.353 \\ 
  14 &  &  & 0.200 & 0.400 & 0.671 & 0.429 \\ 
  15 &  &  & 0.400 & 0.200 & 0.778 & 0.612 \\ 
  16 & 0.200 & 0.100 & 0.100 & 0.100 & 1.006 & 1.005 \\ 
  17 &  &  & 0.200 & 0.200 & 0.852 & 0.563 \\ 
  18 &  &  & 0.400 & 0.400 & 0.609 & 0.301 \\ 
  19 &  &  & 0.200 & 0.400 & 0.596 & 0.296 \\ 
  20 &  &  & 0.400 & 0.200 & 0.904 & 0.893 \\ 
   \hline
\end{tabular}
\caption{\label{table:results3} Simulation specifications  and ARB(M1a)/ARB(M12), ASRB(M1a)/ASRB(M12) for Case 3.}
\end{table}

\begin{table}[h!t!b!p!]
\centering
\begin{tabular}{ccccccc}
& $p_2$  for $\mu_i$  & $\tau_{11}$ for $\mu_i$  & $p_1$   for $\theta_{ij}$  &  $\tau_{21}$ for $\theta_{ij}$  & M1a/M12(ARB)  & M1a/M12(ASRB) \\ 
  \hline
1 & 0.100 & 0.050 & 0.100 & 0.100 & 1.353 & 1.545 \\ 
  2 &  &  &  & 0.200 & 1.175 & 1.344 \\ 
  3 &  &  &  & 0.400 & 0.815 & 0.777 \\ 
  4 &  &  & 0.200 & 0.100 & 1.270 & 1.402 \\ 
  5 &  &  &  & 0.200 & 1.122 & 1.254 \\ 
  6 &  &  &  & 0.400 & 0.775 & 0.807 \\ 
  7 &  & 0.100 & 0.100 & 0.100 & 1.059 & 0.725 \\ 
  8 &  &  &  & 0.200 & 0.865 & 0.584 \\ 
  9 &  &  &  & 0.400 & 0.749 & 0.504 \\ 
  10 &  &  & 0.200 & 0.100 & 1.025 & 0.693 \\ 
  11 &  &  &  & 0.200 & 0.804 & 0.684 \\ 
  12 &  &  &  & 0.400 & 0.760 & 0.817 \\ 
  13 &  & 0.200 & 0.100 & 0.100 & 0.687 & 0.444 \\ 
  14 &  &  &  & 0.200 & 0.651 & 0.298 \\ 
  15 &  &  &  & 0.400 & 0.548 & 0.223 \\ 
  16 &  &  & 0.200 & 0.100 & 0.673 & 0.429 \\ 
  17 &  &  &  & 0.200 & 0.644 & 0.294 \\ 
  18 &  &  &  & 0.400 & 0.582 & 0.424 \\ 
  19 & 0.200 & 0.050 & 0.100 & 0.100 & 1.183 & 1.288 \\ 
  20 &  &  &  & 0.200 & 1.088 & 1.002 \\ 
  21 &  &  &  & 0.400 & 0.875 & 0.679 \\ 
  22 &  &  & 0.200 & 0.100 & 1.216 & 1.076 \\ 
  23 &  &  &  & 0.200 & 1.009 & 1.007 \\ 
  24 &  &  &  & 0.400 & 0.747 & 0.822 \\ 
  25 &  & 0.100 & 0.100 & 0.100 & 0.832 & 0.586 \\ 
  26 &  &  &  & 0.200 & 0.797 & 0.472 \\ 
  27 &  &  &  & 0.400 & 0.710 & 0.426 \\ 
  28 &  &  & 0.200 & 0.100 & 0.903 & 0.637 \\ 
  29 &  &  &  & 0.200 & 0.821 & 0.594 \\ 
  30 &  &  &  & 0.400 & 0.733 & 0.610 \\ 
  31 &  & 0.200 & 0.100 & 0.100 & 0.695 & 0.559 \\ 
  32 &  &  &  & 0.200 & 0.645 & 0.385 \\ 
  33 &  &  &  & 0.400 & 0.502 & 0.160 \\ 
  34 &  &  & 0.200 & 0.100 & 0.679 & 0.450 \\ 
  35 &  &  &  & 0.200 & 0.616 & 0.429 \\ 
  36 &  &  &  & 0.400 & 0.590 & 0.374 \\ 
   \hline
\end{tabular}
\caption{\label{table:results4} Simulation specifications  and ARB(M1a)/ARB(M12), ASRB(M1a)/ASRB(M12)  for Case 4.}
\end{table}
 
Consider Case 1 (both ``outliers'', Table \ref{table:results1}) and ARB. The gains for M1a (vs. M12) increase, as expected, as both $\tau_{11}$'s, $  \tau_{11}(\mu_i) $ for $\mu_i$ and $\tau_{11}(\theta_{ij})$ for $\theta_{ij}$ increase.
The gains for M1a are generally larger for $\left( \tau_{11}(\mu_i),  \tau_{11}(\theta_{ij}) \right)  = (0.10, 0.05)$ than for $(0.05, 0.10).$ That is,  the gains are larger when the variance of $\mu_i$ is larger than the variance of $\theta_{ij}$ in the two outlier models.
The gains for M1a 
decrease as  both $p_2$'s, $p_2(\mu_i)$ and $p_2(\theta_{ij}),$ increase from 0.1 to 0.2 to 0.4. This occurs because
larger $p_2$ in the outlier model makes it closer to M12.
There is no pattern in the gains associated with  $\left( p_2(\mu_i), p_2(\theta_{ij})\right) =(0.1, 0.2)$ vs. those associated with $\left( p_2(\mu_i), p_2(\theta_{ij})\right)= (0.2, 0.1).$
The patterns for ASRB are similar to those for ARB.

Next, consider Case 2 ($\mu_i$ is ``mixture"; $\theta_{ij}$ is ``outlier",  Table \ref{table:results2}) and ARB. The gains for M1a (vs. M12) increase, as expected, as $\tau_{21} (= \tau_{11})$ increases. The gains for M1a are larger for ($\tau_{21}, \tau_{11}) = (0.1, 0.2)$ than for $(0.2,0.1).$ That is, the gains are larger when the variance of $\theta_{ij}$ is larger than the variance of $\mu_i.$ The gains for M1a increase as $p_1 (= p_2)$ increases from 0.1 to 0.2 while there are only small differences for $(p_1, p_2) = (0.1, 0.2)$ vs. $(0.2, 0.1).$ The patterns for ASRB are similar to those for ARB for the specifications with changing variances. For the specifications with changing probabilities the patterns are inconsistent.

Case 3 has  both $\mu_i$ and $\theta_{ij}$ as ``mixtures'' (Table \ref{table:results3}).  For ARB the gains increase as $\tau_{21}$ for $\mu_i$ increases from 0.1 to 0.2 to 0.4 while the changes in the gains are usually small as $\tau_{21}$ for $\theta_{ij}$ increases from 0.1 to 0.2 to 0.4. There are only small changes as $p_1$ for $\mu_i,$  $p_1(\mu_i)$,  increases from 0.1 to 0.2, and no  pattern as $p_1$ for $\theta_{ij},$ $p_1(\theta_{ij})$, increases.  
The patterns for ASRB are similar to those for ARB.

Case 4 has an ``outlier'' model for $\mu_i$ and a ``mixture'' model for $\theta_{ij}$ (Table \ref{table:results4}). For ARB the gains increase as $\tau_{11}$ increases from 0.05 to 0.10 to 0.20 and $\tau_{21}$ increases from 0.1 to 0.2 to 0.4. For almost all specifications there is an increase in the gains as $p_1$ and $p_2$ increase from 0.1 to 0.2.  The patterns for ASRB are similar to those for ARB.

\subsection{Two sources vs. one source}

With data from two sources one could make inference using the data from only one source. Our investigation started with data generated from Cases 1 - 4. For each case and specification we calculated ARB and ASRB using models M1a, M1b, M11a, M11b and M12. We also used MSA and MBR, one source models (Section 3.4) using only the SAHIE and BRFSS data. As in Section 3.2 we summarize for each case using discrepancy ratios, here A(M12)/A(M1a), A(MSA)/A(M1a) and A(MBR)/A(M1a) where A denotes either ARB or ASRB. Tables \ref{table:summ1b}-\ref{table:summ4b} summarize the results for Cases 1 - 4.  For each ratio we give the minimum value, first quartile, median, mean, third quartile and maximum value among the set of specifications for the case. Except for a few minimum values all discrepancy ratios exceed 1, meaning that there are losses by using M12, MSA and MBR relative to M1a. For ARB the median values of the ratio ARB(MBR)/ARB(M1a) are very large, i.e., (2.530, 2.255, 1.782, 2.703) for Cases 1 - 4.
This shows the benefit of using a GL model, here M1a. The median values for ASRB(MBR)/ASRB(M1a) (4.593, 5.319, 3.636, 6.021) are even larger and provide additional evidence about using M1a.  





\begin{table}[h!t!b!p!]
\centering
\begin{tabular}{rrrr}
&   & ARB    &  \\ 
 & M12/M1a & MSA/M1a & MBR/M1a \\ 
  \hline
Min. & 1.141 & 1.089 & 1.458 \\ 
  1st.Qu. & 1.494 & 1.516 & 2.019 \\ 
  Median & 1.722 & 1.875 & 2.530 \\ 
  Mean & 1.996 & 2.044 & 2.659 \\ 
  3rd.Qu. & 2.219 & 2.316 & 3.242 \\ 
  Max & 4.737 & 3.950 & 4.969 \\ 
  &   & &     \\ 
 &   & ASRB    &  \\ 
 & M12/M1a & MSA/M1a & MBR/M1a \\ 
  Min & 1.176 & 1.319 & 1.826 \\ 
  1st.Qu  & 1.786 & 2.153 & 2.790 \\ 
  Median  & 2.235 & 3.934 & 4.593 \\ 
  Mean & 3.329 & 5.741 & 6.526 \\ 
  3rd.Qu & 3.553 & 7.445 & 8.436 \\ 
  Max  & 18.581 & 32.331 & 30.281 \\ 
   \hline
\end{tabular}
\caption{\label{table:summ1b} Summary statistics for ARB and ASRB for Case 1: One vs. two sources. }
\end{table}


\begin{table}[h!t!b!p!]
\centering
\begin{tabular}{rrrr}
&   & ARB    &  \\ 
 & M12/M1a & MSA/M1a & MBR/M1a \\ 
  \hline
 Min & 1.012 & 0.965 & 1.871 \\ 
   1st.Qu  & 1.123 & 1.107 & 2.044 \\ 
  Median & 1.345 & 1.255 & 2.255 \\ 
  Mean & 1.482 & 1.364 & 2.235 \\ 
   3rd.Qu  & 1.947 & 1.620 & 2.419 \\ 
  Max  & 2.113 & 2.054 & 2.608 \\ 
  &   & &     \\ 
 &   & ASRB    &  \\ 
 & M12/M1a & MSA/M1a & MBR/M1a \\ 
  1st.Qu  & 1.247 & 1.639 & 3.568 \\ 
  Median  & 1.650 & 2.520 & 4.716 \\ 
  Mean  & 2.279 & 3.546 & 5.319 \\ 
   3rd.Qu  & 3.147 & 5.855 & 6.963 \\ 
  Max  & 5.246 & 8.317 & 10.078 \\ 
   \hline
\end{tabular}
\caption{\label{table:summ2b}Summary statistics for ARB and ASRB for Case 2: One vs. two sources.}
\end{table}



\begin{table}[h!t!b!p!]
\centering
\begin{tabular}{rrrr}
&   & ARB    &  \\ 
 & M12/M1a & MSA/M1a & MBR/M1a \\ 
  \hline
 Min & 0.943 & 1.370 & 1.445 \\ 
   1st.Qu & 1.155 & 1.551 & 1.594 \\ 
  Median & 1.254 & 1.724 & 1.782 \\ 
  Mean & 1.304 & 1.813 & 1.851 \\ 
   3rd.Qu  & 1.510 & 1.964 & 1.985 \\ 
  Max  & 1.733 & 2.643 & 2.653 \\ 
  &   & &     \\ 
 &   & ASRB    &  \\ 
 & M12/M1a & MSA/M1a & MBR/M1a \\ 
 Min & 0.935 & 1.787 & 2.139 \\ 
   1st.Qu  & 1.081 & 2.198 & 2.395 \\ 
  Median  & 1.881 & 3.509 & 3.636 \\ 
  Mean  & 2.096 & 3.846 & 4.127 \\ 
   3rd.Qu  & 2.321 & 4.307 & 4.556 \\ 
  Max  & 4.867 & 8.459 & 12.665 \\ 
   \hline
\end{tabular}
\caption{\label{table:summ3b} Summary statistics for ARB and ASRB for Case 3: One vs. two sources.}
\end{table}


\begin{table}[h!t!b!p!]
\centering
\begin{tabular}{rrrr}
&   & ARB    &  \\ 
 & M12/M1a & MSA/M1a & MBR/M1a \\ 
  \hline
 Min  & 0.726 & 1.681 & 1.590 \\ 
   1st.Qu  & 1.019 & 2.400 & 2.317 \\ 
  Median & 1.260 & 2.879 & 2.703 \\ 
  Mean & 1.269 & 3.150 & 2.904 \\ 
   3rd.Qu & 1.382 & 3.564 & 3.316 \\ 
  Max  & 2.116 & 6.413 & 5.740 \\ 

  &   & &     \\ 
 &   & ASRB    &  \\ 
 & M12/M1a & MSA/M1a & MBR/M1a \\ 
   Min & 0.530 & 1.326 & 1.154 \\ 
   1st.Qu & 1.002 & 3.203 & 3.393 \\ 
  Median   & 1.407 & 6.507 & 6.021 \\ 
  Mean  & 1.817 & 11.151 & 10.253 \\ 
   3rd.Qu  & 2.040 & 11.854 & 12.164 \\ 
  Max  & 5.935 & 56.195 & 46.653 \\ 
   \hline
\end{tabular}
\caption{\label{table:summ4b} Summary statistics for ARB and ASRB for Case 4: One vs. two sources. }
\end{table}

Similarly, the four distributions of A(MSA)/A(M1a) have large values, albeit usually smaller than for A(MBR)/A(M1a). For example, for ARB the medians of \{ARB(MSA)/ARB(M1a), ARB(MBR)/ARB(M1a)\} are \{(1.875, 2.530), (1.255, 2.255), (1.724, 1.782), (2.879, 2.703)\}.

We next refined the specifications in Cases 1 - 4 by assuming different distributions for the $\theta'$s. This is more realistic in that, typically, the two sources would be expected to have different characteristics. We also simplified by assuming no aberrant values at the county level. Throughout we generate data with  Source 1 as BRFSS and Source 2 as SAHIE. For Case 5 the model used to generate the data is 

$$y_{ij} \stackrel {ind}{\sim} N(\theta_{ij}, V_{ij})$$
$$\theta_{i1} \stackrel {ind}{\sim} N(\mu_i, \tau^2_1)$$
$$\theta_{i2} \stackrel {ind}{\sim} N(\mu_i, \delta \tau^2_2)$$
where  $\delta \in \{0,1\}$ and  $P(\delta = 1) = p.$  Finally, $\mu_i \stackrel {iid}{\sim} N(0.25, \tau^2).$

We used 18 specifications which included all arrangements of $\tau = 0.05, \tau_1 = (0.005, 0.010), p = (0.10, 0.20, 0.40)$ and $\tau_2 = (0.05, 0.10, 0.20)$: see Table  \ref{table:specs6}.  The results, summarized in Table \ref{table:summ6}, show almost no specifications where using the BRFSS data as a single source is preferable to using M1a. For ARB and ASRB the median values of A(MBR)/A(M1a) (across the 18 specifications) are quite large, i.e., 1.663 and 2.070. (Note that another study, using larger values of $\tau$, produced very similar results.) Looking at the 18 individual ratios, A(MBR)/A(M1a), the values of the ratios decrease as $p$ and $\tau_2$ increase, as expected. The value of $\tau_1$ has only a minimal effect. See Tables \ref{table:specs6}, \ref{table:arb6} and \ref{table:asrb6}.  


\begin{table}[h!t!b!p!]
\centering
\begin{tabular}{rrrrr}
 & $\tau$ & $\tau_{1}$ & $p$ & $\tau_{2}$  \\ 
  \hline
 1 & 0.05  & 0.005 & 0.1 & 0.05  \\ 
  2 &  &  &  & 0.1 \\ 
  3 &  &  &  & 0.2 \\ 
  4 &  &  & 0.2  & 0.05  \\ 
  5 &  &  &  & 0.1 \\ 
  6 &  &  &  & 0.2 \\ 
  7 &  &  & 0.4  & 0.05  \\ 
  8 &  &  &  & 0.1 \\ 
  9 &  &  &  & 0.2  \\ 
  10 &  & 0.01  & 0.1  & 0.05  \\ 
  11 &  &  &  & 0.1  \\ 
  12 &  &  &  & 0.2  \\ 
  13 &  &  & 0.2  & 0.05  \\ 
  14 &  &  &  & 0.1 \\ 
  15 &  &  &  & 0.2  \\ 
  16 &  &  & 0.4  & 0.05  \\ 
  17 &  &  &  & 0.1  \\ 
  18 &  &  &  & 0.2 \\ 
   \hline
\end{tabular}
\caption{\label{table:specs6} Simulation specifications for Case 5: $\mu_i\sim N\left(0.25, \tau^2 \right), \theta_{i1} \sim N \left( \mu_i, \tau_{1}^2 \right), \theta_{i2} \sim N \left( \mu_i,  \delta \tau_{2}^2 \right), \delta \sim Ber(p).$}
\end{table}

\begin{table}[h!t!b!p!]
\centering
\begin{tabular}{rrrr}
&   & ARB    &  \\ 
 & M12/M1a & MSA/M1a & MBR/M1a \\ 
  \hline
Min  & 1.000 & 1.080 & 1.088 \\ 
   1st.Qu  & 1.065 & 1.346 & 1.443 \\ 
  Median & 1.162 & 1.485 & 1.663 \\ 
  Mean & 1.285 & 1.786 & 1.694 \\ 
  3rd.Qu  & 1.326 & 2.002 & 2.025 \\ 
  Max  & 2.030 & 3.435 & 2.342 \\ 
  
  &   & &     \\ 
 &   & ASRB    &  \\ 
 & M12/M1a & MSA/M1a & MBR/M1a \\ 
 Min  & 1.000 & 1.468 & 0.962 \\ 
   1st.Qu  & 1.133 & 2.233 & 1.712 \\ 
  Median  & 1.348 & 4.369 & 2.070 \\ 
  Mean  & 1.681 & 7.033 & 2.377 \\ 
   3rd.Qu  & 1.851 & 11.494 & 3.112 \\ 
  Max  & 3.899 & 20.498 & 4.465 \\ 
   \hline
\end{tabular}
\caption{\label{table:summ6} Summary statistics  for ARB and ASRB for Case 5.}
\end{table}

For Case 6  the model used to generate the data is the same as that for Case 5 except that $p = 1.$ Since Case 6 assumes no aberrations a global-local model is unnecessary. Thus, it is not surprising that there are specifications where MBR is preferred. For example, the medians of A(MBR)/A(M1a) over the 12 specifications are 0.854 and 0.731 for ARB and ASRB (see Table \ref{table:summ5}). The 12 specifications are listed in Table \ref{table:specs5}.
As seen in Tables \ref{table:arb5} and \ref{table:asrb5} 
the value of $\tau_1$ has little effect. However, as $\tau_2$ increases    A(MBR)/A(M1a) decreases while A(MSA)/A(M1a) increases. Thus, for large values of $\tau_2$ there are benefits to using MBR rather than the GL model M1a.  But, as seen in Case 5, it seems unlikely that using only one source of data will be preferable when (a)there are significant aberrant observations, and (b)the single source is the one with small bias and large sampling variance.



\begin{table}[h!t!b!p!]
\centering
\begin{tabular}{rrrr}
 & $\tau$ & $\tau_{1}$ & $\tau_{2}$  \\ 
  \hline
1 & 0.05 & 0.005 & 0.01 \\ 
  2 &  &  & 0.02 \\ 
  3 &  &  & 0.05 \\ 
  4 &  &  & 0.1 \\ 
  5 &  &  & 0.2 \\ 
  6 &  &  & 0.4 \\ 
  7 &  & 0.01 & 0.01 \\ 
  8 &  &  & 0.02 \\ 
  9 &  &  & 0.05 \\ 
  10 &  &  & 0.1 \\ 
  11 &  &  & 0.2 \\ 
  12 &  &  & 0.4 \\ 
   \hline
\end{tabular}
\caption{\label{table:specs5}  Simulation specifications for Case 6: $\mu_i\sim N\left(0.25, \tau^2 \right), \theta_{ij} \sim N \left( \mu_i, \tau_{j}^2 \right).$ }
\end{table}

\begin{table}[h!t!b!p!]
\centering
\begin{tabular}{rrrr}
&   & ARB    &  \\ 
 & M12/M1a & MSA/M1a & MBR/M1a \\ 
  \hline
Min  & 0.954 & 1.065 & 0.546 \\ 
  1st.Qu  & 0.967 & 1.208 & 0.634 \\ 
  Median & 1.000 & 2.293 & 0.854 \\ 
  Mean & 1.027 & 3.283 & 1.097 \\ 
  3rd.Qu & 1.099 & 4.596 & 1.495 \\ 
  Max  & 1.170 & 8.230 & 2.168 \\ 
   &   & &     \\ 
 &   & ASRB    &  \\ 
 & M12/M1a & MSA/M1a & MBR/M1a \\ 
Min & 0.882 & 1.083 & 0.293 \\ 
  1st.Qu  & 0.911 & 1.452 & 0.380 \\ 
  Median  & 0.975 & 5.067 & 0.731 \\ 
  Mean  & 1.030 & 15.203 & 1.547 \\ 
  3rd.Qu  & 1.182 & 18.322 & 2.420 \\ 
  Max & 1.289 & 60.459 & 4.745 \\ 
   \hline
\end{tabular}
\caption{\label{table:summ5} Summary statistics  for ARB and ASRB for Case 6.}
\end{table}

\section{Discussion and Summary}

In the context of small area estimation from a single survey, \cite{dl95} developed a hierarchical Bayes method using
a general class of scale mixture of normal distributions (see Model 1 on page 314). For certain choices of prior their
hierarchical Bayes methodology ensures robustness against outliers (see their Theorem 5). In the context of a single
survey the horseshoe prior is included in the class of priors proposed in \cite{dl95}, so protects against outliers
in the sense of their Theorem 5. Extension to the several survey case, as in this paper, would be valuable.

\bigskip

A referee asked about formal model comparisons. For the model in (\ref{FH model}) and (\ref{One level shrinkage prior})
(with $\eta$ as a linear regression) \cite{tghs18} consider choosing a single model using DIC, and find in their simulation
study that ``the selected model produces deviation measures close to the smallest among all candidate models and better
coverage rate than the DM model." However, there has been extensive criticism of DIC, and one should be cautious
about using it for models such as (\ref{FH model}) and (\ref{One level shrinkage prior}) and for the more complex models
that we have used. \cite{p08} is a thorough investigation of DIC that details several significant problems. \cite{cl09} identify additional problems with DIC, and also describe (Section 4.6.2) another general approach, i.e.,
predictive model selection.

\bigskip

It is of interest to investigate the effect of having a larger number of sources of information. We did a preliminary study using the four cases that are the basis of our simulation study. To illustrate, consider the 30 specifications for Case 1 in Table \ref{table:results1} . We generated data as in Section 3 for $J = 2$ and $J = 4$ sources. The $V_{ij}$ were selected as a bootstrap sample from the $\hat{V}_{ij}$. There are, as expected, substantial gains as measured using the ratios (A(Mx($J = 4$))/A(Mx($J = 2$)) where A denotes ARB or ASRB. But, they vary considerably over the choices of the probabilities and variances. For example, for M1a the smallest and largest values of the ratio are 0.142 and 0.767. Consider M1a and ARB: as both $\tau_{11}$'s increase the gains from using $J = 4$ sources decrease. As both $p_2$'s increase from 0.10 to 0.20 to 0.40 the dominant pattern is for the ratio to decrease, then increase. Further study is needed to get more definitive results.

Making inference about the sampling variances, the $V_{ij}$, is a challenging problem. The most common method is to assume a model for the $\hat{V}_{ij}$, the estimates of the $V_{ij}$. However, with only a single sample, $\{\hat{V}_{ij}: j = 1,...,J; i = 1,...,I\}$, one cannot verify the model. Methodology proposed in the literature is reviewed by \cite{cas23}. In the GL setting, but with only one data set, \cite{tghs18}  assume fixed $V_i$ to avoid problems with identifiability. Our situation is more complex, i.e., with small areas and several data sources.
\\

As future research we envision modelling the $\{\hat{V}_{ij}: j = 1,...,J; i = 1,...,I\}$ using data sets from prior years.
An additional extension is to include covariates in our models.
\\

The objective of this paper is to present methodology to improve inference for ``small area'' parameters by using data from several sources. While there is an extensive literature about global-local models there appears to be only one other paper, \cite{sag23}, that uses a several stage model. Moreover, the application in \cite{sag23} is completely different from ours. Consequently, it was necessary to supplement our analysis of the Florida county data with an extensive simulation study to show properties of this methodology. 

We introduce a set of global-local models and explain how these models combine information across counties and sources. Analyzing the BRFSS and SAHIE data using these models provides an opportunity to show how these models treat aberrant observations. This also includes a comparison of inferences using all of the data with inferences based on only one data source. In the extensive simulation study the data are selected from both outlier and mixture models. These results show that the proposed methodology can be used to provide improved inferences for small area parameters by using several data sources.

\section{Acknowledgment}
\emph{The authors are grateful to Advanced Cyberinfrastructure Coordination Ecosystem: Services \& Support (ACCESS) for the computing support, and to several reviewers who posed interesting questions, valuable for future research.}

\clearpage 
\section{Appendix}

\renewcommand{\thefigure}{\arabic{figure}A} 
\setcounter{figure}{0}


\subsection{The $\phi_i$'s \label{phis}}

\begin{figure}[h!t!b!p!]
\hspace{-0.25in}
\centering
 \includegraphics[height=3.75in, width=7in]{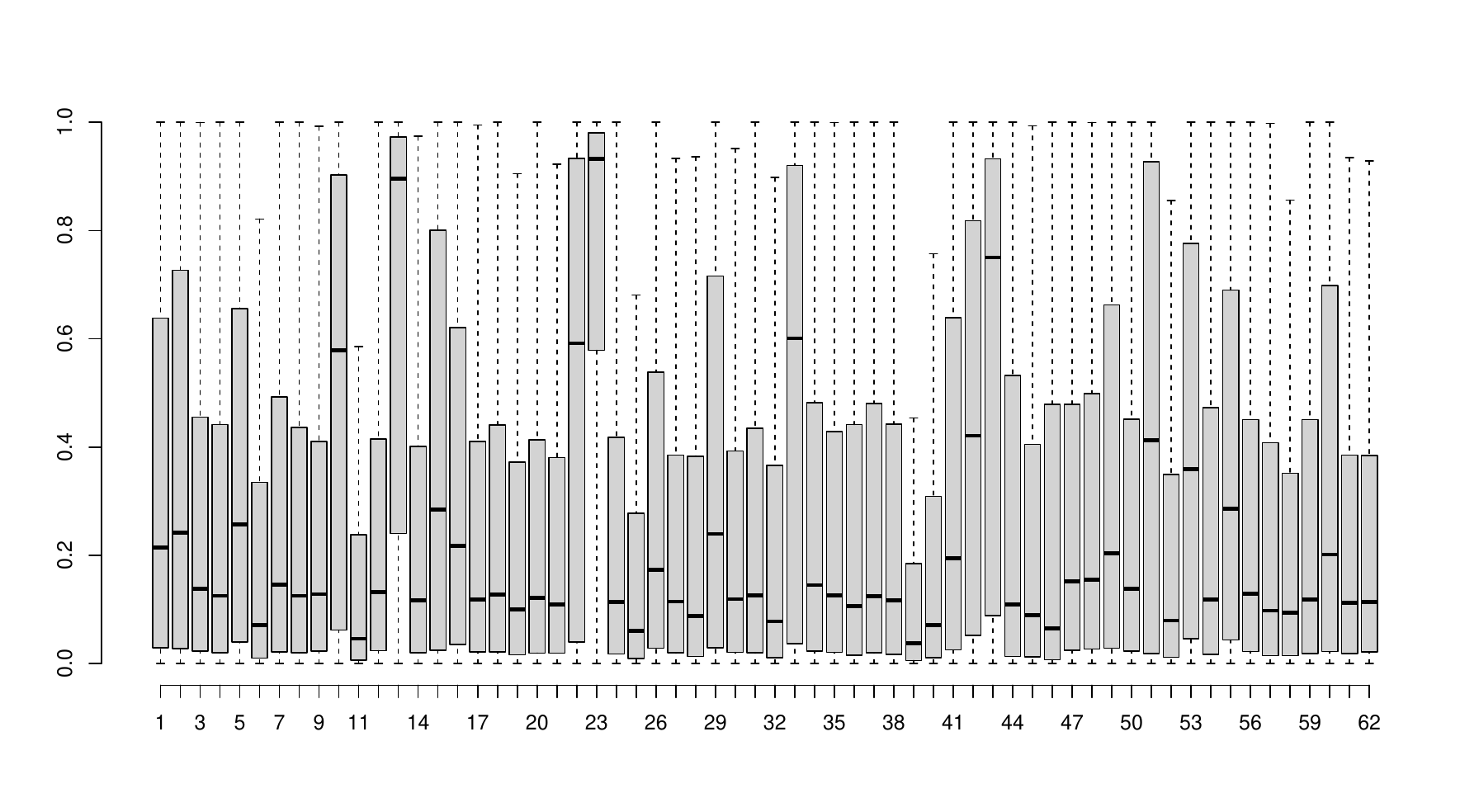} 
 \includegraphics[height=3.75in, width=7in]{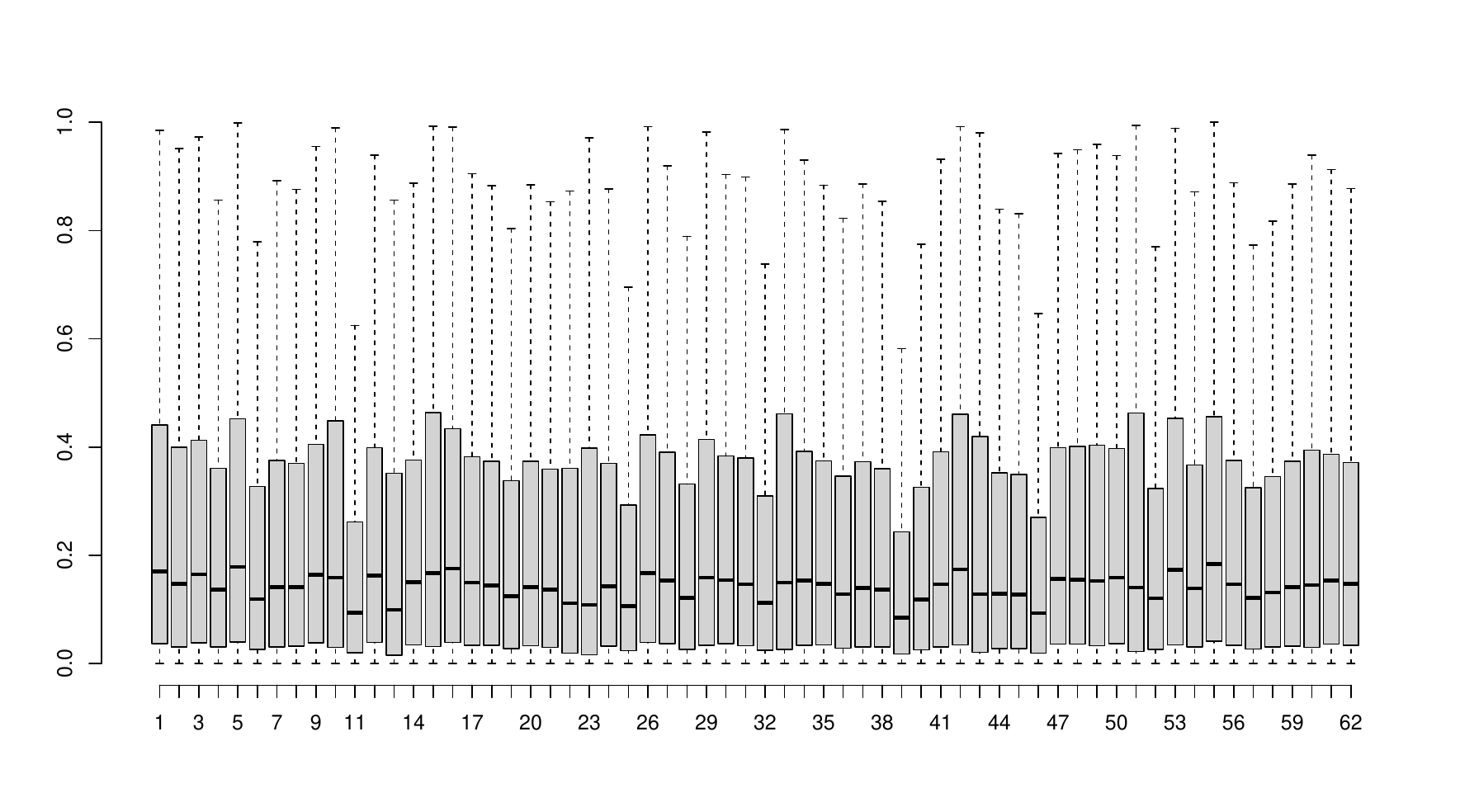} 

\caption{\label{fig:M1sphi}  The $\phi_i$'s for M1a (top) and M1b (bottom).}
 \end{figure}

\begin{figure}[h!t!b!p!]
\hspace{-0.25in}
\centering
\includegraphics[height=3.75in, width=7in]{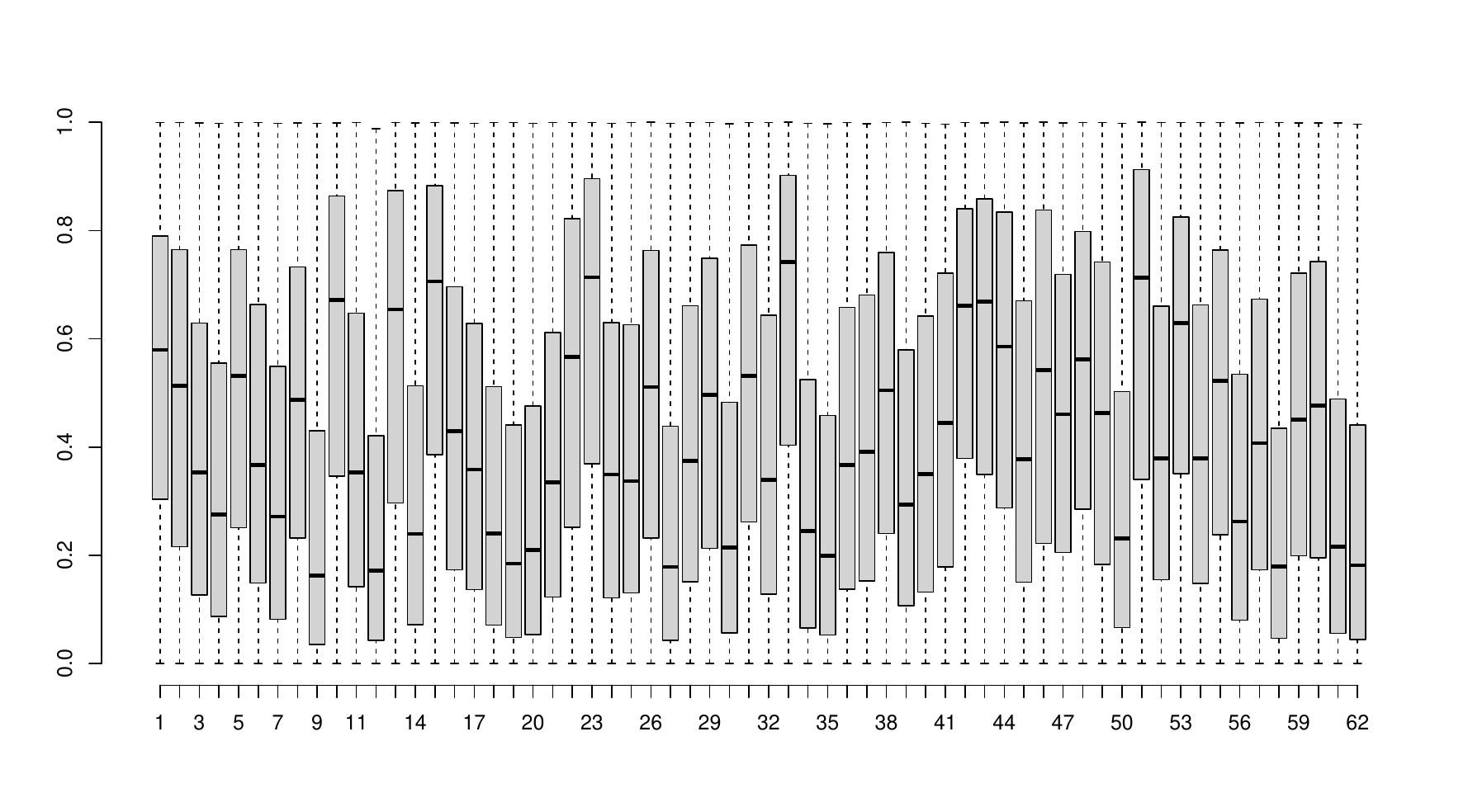} 
 \includegraphics[height=3.75in, width=7in]{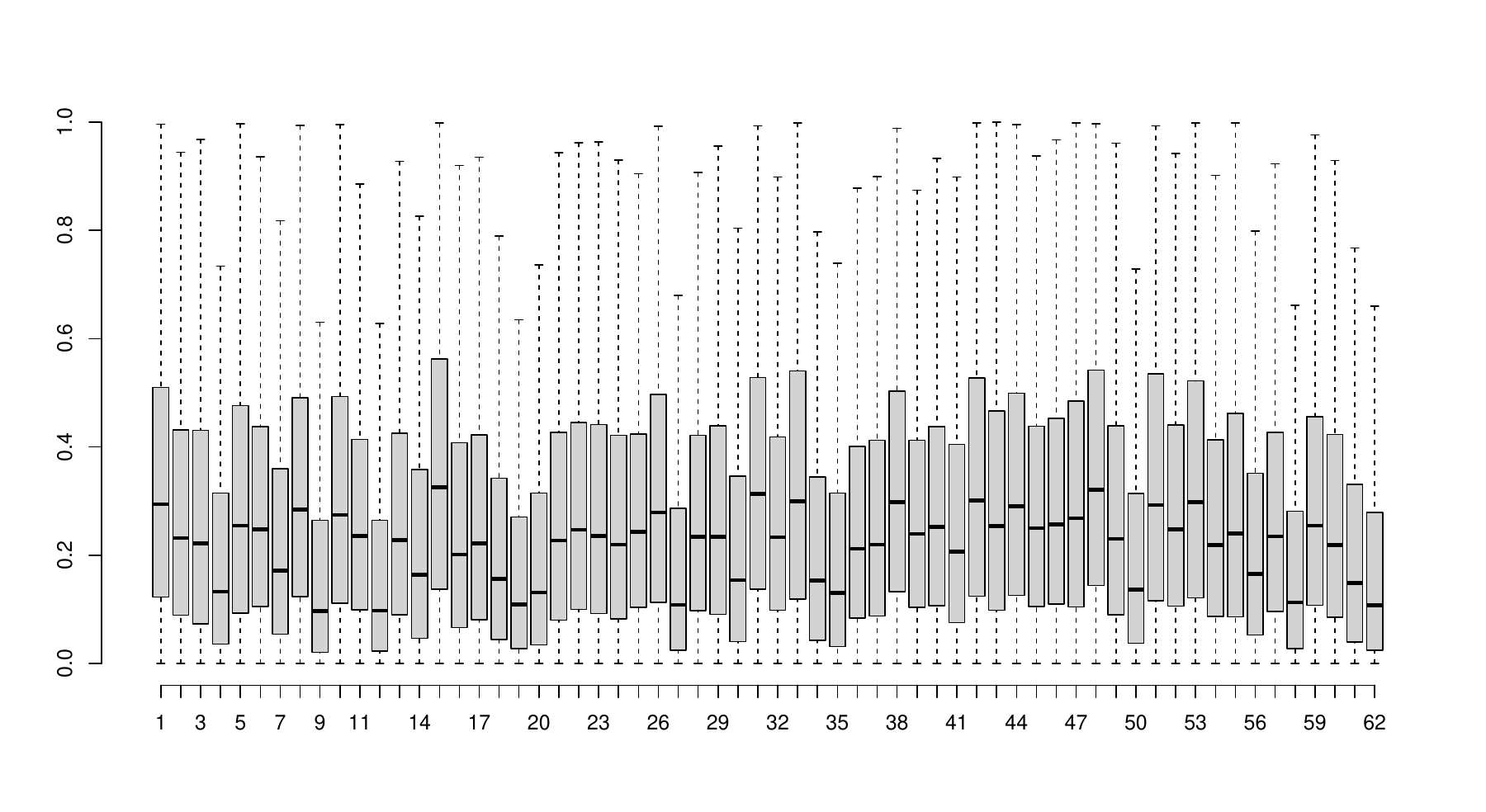} 
\caption{\label{fig:M11sphi}  The $\phi_i$'s for M11a (top) and M11b (bottom).}
 \end{figure}

\begin{figure}[h!t!b!p!]
\hspace{-0.25in}
\centering
 \includegraphics[height=3.75in, width=7in]{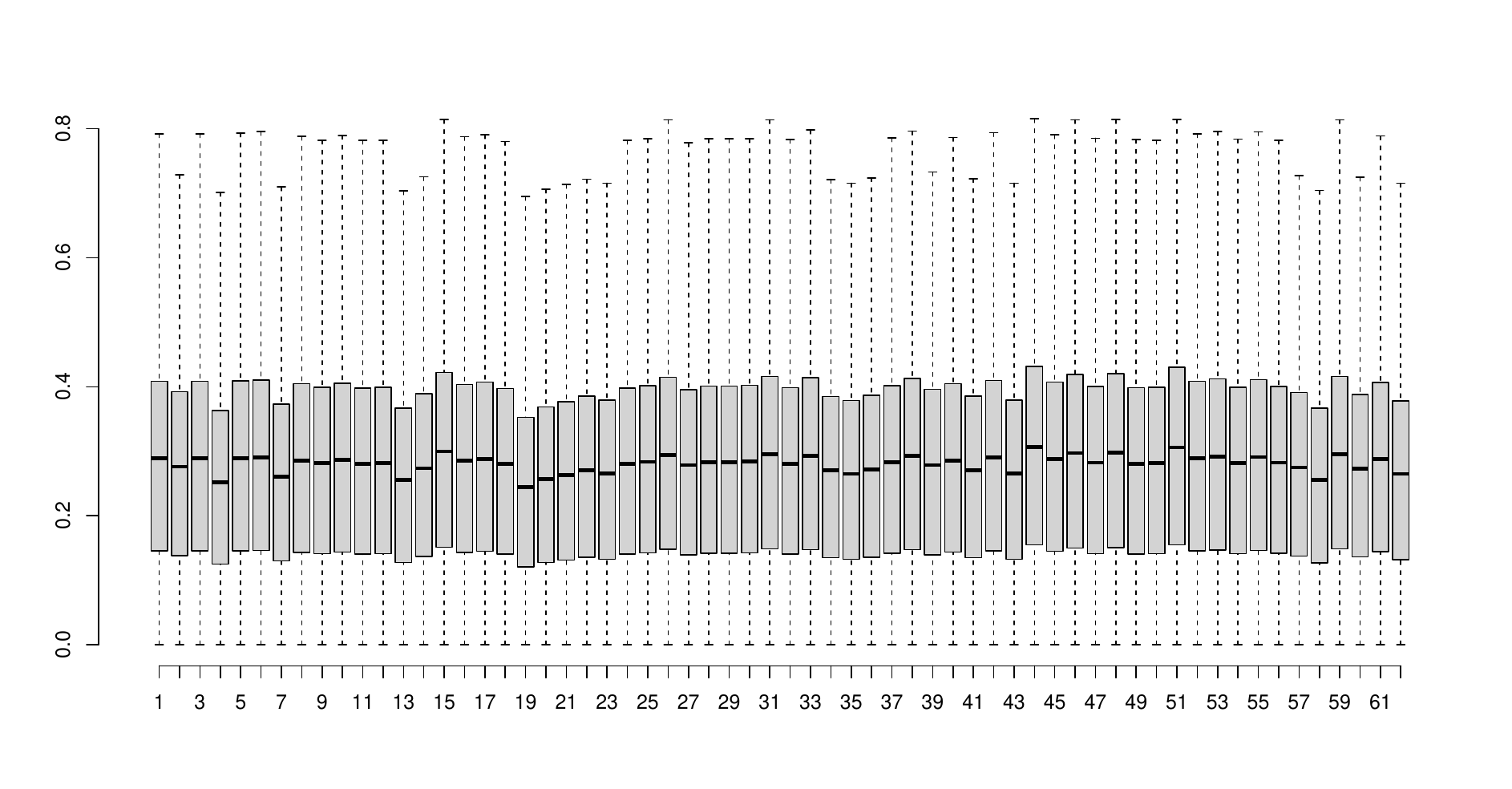} 
\caption{\label{fig:M12phi}  The $\phi_i$'s for  M12.}
 \end{figure}

\subsection{Full conditional distributions for M11a and M11b  \label{cond}}

We present below the full conditional distributions for the interaction effects, $\theta_{ij},$ the small area expected values, $\mu_i,$ and the local and global variances.

\emph{Interaction effects:}
 $\theta_{ij} \; | \; else  \; \sim \; N( \overline{\theta}_{ij}, V_{\theta_{ij}} )$ where
\[
\overline{\theta}_{ij} = \frac{\frac{Y_{ij}}{V_{ij}}+\frac{\mu_i}{\lambda^2_{ij}\lambda_i^2\tau_1^2}}{\frac{1}{V_{ij}}+ \frac{1}{\lambda^2_{ij}\lambda_i^2\tau_1^2}} \qquad\text{and} \qquad  V_{\theta_{ij}} = \frac{1}{\frac{1}{V_{ij}} + \frac{1}{\lambda^2_{ij}\lambda_i^2\tau_1^2}}.
\]
\emph{Small area means:}
$\mu_i \; |\; else \; ~\sim \;   N( \overline{\mu}_i, V_{\mu_i} )$ where
\[
 \overline{\mu}_i, = \frac{\overline{\theta}_i c_i + \eta d_i }{c_i+ d_i},  \qquad \qquad  V_{\mu_i} = \frac{1}{c_i+d_i},
\]
$a_{ij}= \lambda^2_{ij}\lambda_i^2\tau_1^2,  c_i=\sum_j (1/a_{ij}), d_i = 1/(\lambda_i^2\tau_2^2),$ and $\overline{\theta}_i  = \frac{\sum_j \left(\theta_{ij}/a_{ij}\right) }{\sum_j \left(1/a_{ij}\right)}.$
\\

\emph{Local variances:} Following \cite{mas16}, we use the following Inverse Gamma ( $IG(\text{shape}, \text{rate})$ ) scale mixture representation:  If $\kappa \; | \; \xi \; \sim \; IG(1/2, 1/\xi)$ and  $\xi \; \sim   \; IG(1/2, 1 )$ then   $\kappa \sim  \mathcal{C}^+ (0,1)$ and $\nu=\kappa^2$ (see Section 2) has  the horseshoe distribution.  Note that the $IG(\text{shape, rate})$  density is 
proportional to $x^{(-1-\text{shape})} e^{(-\text{rate}/x)}.$ 

Using the horseshoe prior on the local variances,

\[
\lambda_{ij}^2 \; | \; else \; \sim \; IG\left(1, \frac{(\theta_{ij}-\mu_i)^2 }{2\lambda_i^2 \tau_1^2} + \frac{1}{\xi} \right); 
  \qquad  \xi  \; | \;  \lambda_{ij}^2   \; \sim \; IG\left(1, 1 + \frac{1}{\lambda_{ij}^2} \right).
\] 

\[
\lambda_i^2 \; | \; else \; \sim \; IG\left( \frac{(J+4)}{2} -1,  \sum_j \frac{(\theta_{ij}-\mu_i)^2 }{2\lambda_{ij}^2 \tau_1^2} +  \frac{(\mu_i-\eta)^2 }{2\tau_2^2} +  \frac{1}{\xi} \right); 
  \qquad  \xi  \; | \;  \lambda_i^2   \; \sim \; IG\left(1, 1 + \frac{1}{\lambda_i^2} \right).
\]

Using the \emph{lasso} prior $(\propto \exp(-\kappa^2)$) on the local variances, and the Generalized Inverse Gaussian  distribution $GIG(\text{shape}, \chi, \psi) \propto   x^{(\text{shape}-1)} e^{-\frac{1}{2} ( \chi/x + \psi x)},$ 

\[
\lambda_{ij}^2 \; | \; else \; \sim \; GIG\left(1/2, \frac{(\theta_{ij}-\mu_i)^2 }{\lambda_i^2 \tau_1^2}, 2 \right)
\]
and
\[
 \lambda_i^2 \; | \; else \; \sim \; GIG\left( \frac{(-J+1)}{2} ,  \sum_j \frac{(\theta_{ij}-\mu_i)^2 }{\lambda_{ij}^2 \tau_1^2} +  \frac{(\mu_i-\eta)^2 }{\tau_2^2} ,2 \right).
\]


\emph{Global variances:} Using the horseshoe prior on the global variances, we have the following scale mixture representation:
 
\[
\tau_1^2 \; | \; else \; \sim \; IG\left( \frac{(IJ+3)}{2} - 1, \frac{(\theta_{ij}-\mu_i)^2 }{2\lambda_{ij}^2 \lambda_i^2} + \frac{1}{\xi} \right); 
  \qquad  \xi  \; | \;  \tau_1^2   \; \sim \; IG\left(1, 1 + \frac{1}{\tau_1^2} \right).
\] 

\[
\tau_2^2 \; | \; else \; \sim \; IG\left( \frac{(I+3)}{2} - 1,  \sum_j \frac{(\theta_{ij}-\mu_i)^2 }{2\lambda_i^2}  +  \frac{1}{\xi} \right); 
  \qquad  \xi  \; | \;  \tau_2^2   \; \sim \; IG\left(1, 1 + \frac{1}{\tau_2^2} \right).
\] 


\section{References}
\renewcommand*{\refname}{}

\clearpage
\section{Supplementary Material}

\setcounter{table}{0}
\renewcommand{\thetable}{S\arabic{table}}
\renewcommand*{\theHtable}{\thetable}



\begin{table}[ht]
\centering
\begin{tabular}{lccccc}
  &   & ARB &   &  \\ 
  & M11a/M12 & M11b/M12 & M1a/M12 & M1b/M12 \\ 
  \hline
Min & 0.388 & 0.647 & 0.244 & 0.567 \\ 
  1st.Q & 0.563 & 0.791 & 0.426 & 0.764 \\ 
  Median & 0.633 & 0.853 & 0.589 & 0.802 \\ 
  Mean & 0.658 & 0.839 & 0.572 & 0.803 \\ 
  3rd.Q & 0.764 & 0.905 & 0.705 & 0.868 \\ 
  Max & 0.944 & 0.979 & 0.923 & 0.955 \\ 
  &   & &   &  \\ 
  &   & ASRB &   &  \\ 
  & M11a/M12 & M11b/M12 & M1a/M12 & M1b/M12 \\ 
  \hline
  Min & 0.191 & 0.782 & 0.058 & 0.222 \\ 
  1st.Q & 0.465 & 0.857 & 0.301 & 0.773 \\ 
  Median & 0.546 & 0.913 & 0.474 & 0.926 \\ 
  Mean & 0.592 & 1.130 & 0.470 & 0.960 \\ 
  3rd.Q & 0.699 & 1.162 & 0.653 & 1.066 \\ 
  Max & 1.254 & 3.015 & 0.888 & 2.834 \\ 
   \hline
\end{tabular}
\caption{\label{table:summ1} Summary statistics  for ARB and ASRB for Case 1.}
\end{table}

 \begin{table}[ht]
\centering
\begin{tabular}{lcccc}
  &   & ARB &   &  \\ 
  & M11a/M12 & M11b/M12 & M1a/M12 & M1b/M12 \\ 
  \hline
Min & 0.531 & 0.697 & 0.465 & 0.658 \\ 
  1st.Q & 0.617 & 0.819 & 0.544 & 0.768 \\ 
  Median & 0.760 & 0.874 & 0.733 & 0.851 \\ 
  Mean & 0.754 & 0.866 & 0.724 & 0.853 \\ 
  3rd.Q & 0.894 & 0.938 & 0.900 & 0.949 \\ 
  Max & 0.951 & 0.981 & 0.959 & 0.995 \\ 
  &   & &   &  \\ 
  &   & ASRB &   &  \\ 
  & M11a/M12 & M11b/M12 & M1a/M12 & M1b/M12 \\ 
  \hline
  Min & 0.250 & 0.443 & 0.166 & 0.345 \\ 
  1st.Q & 0.442 & 0.684 & 0.364 & 0.598 \\ 
  Median & 0.607 & 0.790 & 0.575 & 0.685 \\ 
  Mean & 0.607 & 0.781 & 0.565 & 0.705 \\ 
  3rd.Q & 0.827 & 0.925 & 0.799 & 0.889 \\ 
  Max& 0.892 & 1.016 & 0.940 & 0.981 \\ 
   \hline
\end{tabular}
\caption{\label{table:summ2} Summary statistics  for ARB and ASRB for Case 2.}
\end{table}

\begin{table}[h!t!b!p!]
\centering
\begin{tabular}{lcccc}
  &   & ARB &   &  \\ 
  & M11a/M12 & M11b/M12 & M1a/M12 & M1b/M12 \\ 
  \hline
Min & 0.615 & 0.754 & 0.596 & 0.749 \\ 
  1st.Q & 0.688 & 0.823 & 0.650 & 0.840 \\ 
  Median & 0.821 & 0.915 & 0.780 & 0.870 \\ 
  Mean & 0.806 & 0.892 & 0.785 & 0.877 \\ 
  3rd.Q & 0.879 & 0.951 & 0.865 & 0.931 \\ 
  Max & 1.031 & 1.003 & 1.050 & 0.993 \\ 
   &   & &   &  \\ 
 &   & ASRB &   &  \\ 
  & M11a/M12 & M11b/M12 & M1a/M12 & M1b/M12 \\ 
  \hline
  Min & 0.232 & 0.462 & 0.214 & 0.499 \\ 
  1st.Qu & 0.358 & 0.675 & 0.341 & 0.642 \\ 
  Median & 0.492 & 0.754 & 0.498 & 0.792 \\ 
  Mean & 0.554 & 0.761 & 0.575 & 0.781 \\ 
  3rd.Q & 0.728 & 0.906 & 0.796 & 0.930 \\ 
  Max & 1.034 & 0.992 & 1.102 & 1.112 \\ 
   \hline
\end{tabular}
\caption{\label{table:summ3} Summary statistics  for ARB and ASRB for Case 3.}
\end{table}

\begin{table}[h!t!b!p!]
\centering
\begin{tabular}{lcccc}
  &   & ARB &   &  \\ 
  & M11a/M12 & M11b/M12 & M1a/M12 & M1b/M12 \\ 
  \hline
Min & 0.570 & 0.587 & 0.502 & 0.489 \\ 
  1st.Q & 0.807 & 0.835 & 0.677 & 0.775 \\ 
  Median & 0.913 & 0.929 & 0.786 & 0.835 \\ 
  Mean & 1.002 & 0.933 & 0.839 & 0.830 \\ 
  3rd.Q & 1.196 & 1.037 & 1.013 & 0.897 \\ 
  Max & 1.747 & 1.321 & 1.353 & 1.095 \\ 
   &   & &   &  \\ 
 &   & ASRB &   &  \\ 
  & M11a/M12 & M11b/M12 & M1a/M12 & M1b/M12 \\ 
  \hline
Min & 0.205 & 0.389 & 0.160 & 0.318 \\ 
  1st.Qu & 0.551 & 0.798 & 0.429 & 0.721 \\ 
  Median & 0.667 & 0.930 & 0.602 & 0.842 \\ 
  Mean & 0.828 & 1.013 & 0.689 & 0.872 \\ 
  3rd.Q & 0.989 & 1.162 & 0.818 & 1.018 \\ 
  Max & 2.032 & 1.944 & 1.545 & 1.538 \\ 
   \hline
\end{tabular}
\caption{\label{table:summ4} Summary statistics  for ARB and ASRB for Case 4.}
\end{table}

 

\begin{table}[h!t!b!p!]
\centering
\begin{tabular}{ccccc}
 & $p_2$   for $\mu_i$ & $p_2$  for $\theta_{ij}$  & $\tau_{11}$ for $\mu_i$ & $\tau_{11}$ for $\theta_{ij}$ \\
  \hline
1 & 0.1 & 0.1 & 0.025 & 0.025 \\ 
  2 &  &  & 0.05 & 0.05 \\ 
  3 &  &  & 0.1 & 0.1 \\ 
  4 &  &  & 0.2& 0.2 \\ 
  5 &  &  & 0.05 & 0.1 \\ 
  6 &  &  & 0.1 & 0.05 \\ 
  7 & 0.2 & 0.2 & 0.025 & 0.025 \\ 
  8 &  &  & 0.05 & 0.05\\ 
  9 &  &  & 0.1& 0.1 \\ 
  10 &  &  & 0.2& 0.2\\ 
  11 &  &  & 0.05 & 0.1 \\ 
  12 &  &  & 0.1& 0.05 \\ 
  13 & 0.4 & 0.4 & 0.025 & 0.025 \\ 
  14 &  &  & 0.05 & 0.05 \\ 
  15 &  &  & 0.1 & 0.1\\ 
  16 &  &  & 0.2 & 0.2 \\ 
  17 &  &  & 0.05 & 0.1 \\ 
  18 &  &  & 0.1& 0.05 \\ 
  19 & 0.1 & 0.2 & 0.025 & 0.025 \\ 
  20 &  &  & 0.05 & 0.05\\ 
  21 &  &  & 0.1 & 0.1\\ 
  22 &  &  & 0.2& 0.2 \\ 
  23 &  &  & 0.05 & 0.1\\ 
  24 &  &  & 0.1& 0.05 \\ 
  25 & 0.2 & 0.1 & 0.025 & 0.025 \\ 
  26 &  &  & 0.05 & 0.05 \\ 
  27 &  &  & 0.1 & 0.1 \\ 
  28 &  &  & 0.2 & 0.2 \\ 
  29 &  &  & 0.05 & 0.1 \\ 
  30 &  &  & 0.1 & 0.05 \\ 
   \hline
\end{tabular}
\caption{Simulation  specifications for Case 1. } 
\label{table:specs1}  
\end{table}

\begin{table}[h!t!b!p!]
\centering
\begin{tabular}{ccccc}
 & M11a/M12 & M11b/M12 & M1a/M12 & M1b/M12 \\ 
  \hline
1 & 0.825 & 0.867 & 0.803 & 0.874 \\ 
  2 & 0.626 & 0.751 & 0.563 & 0.803 \\ 
  3 & 0.493 & 0.777 & 0.345 & 0.728 \\ 
  4 & 0.388 & 0.961 & 0.244 & 0.771 \\ 
  5 & 0.577 & 0.672 & 0.487 & 0.705 \\ 
  6 & 0.507 & 0.790 & 0.425 & 0.764 \\ 
  7 & 0.858 & 0.882 & 0.828 & 0.931 \\ 
  8 & 0.695 & 0.798 & 0.635 & 0.807 \\ 
  9 & 0.563 & 0.853 & 0.429 & 0.783 \\ 
  10 & 0.505 & 0.930 & 0.298 & 0.771 \\ 
  11 & 0.721 & 0.824 & 0.636 & 0.849 \\ 
  12 & 0.636 & 0.854 & 0.562 & 0.772 \\ 
  13 & 0.929 & 0.928 & 0.896 & 0.955 \\ 
  14 & 0.809 & 0.892 & 0.772 & 0.893 \\ 
  15 & 0.711 & 0.915 & 0.636 & 0.844 \\ 
  16 & 0.674 & 0.979 & 0.612 & 0.910 \\ 
  17 & 0.817 & 0.894 & 0.780 & 0.911 \\ 
  18 & 0.778 & 0.873 & 0.728 & 0.821 \\ 
  19 & 0.944 & 0.908 & 0.923 & 0.939 \\ 
  20 & 0.673 & 0.716 & 0.595 & 0.764 \\ 
  21 & 0.561 & 0.680 & 0.405 & 0.677 \\ 
  22 & 0.619 & 0.967 & 0.255 & 0.817 \\ 
  23 & 0.619 & 0.647 & 0.609 & 0.642 \\ 
  24 & 0.564 & 0.795 & 0.456 & 0.747 \\ 
  25 & 0.814 & 0.891 & 0.796 & 0.938 \\ 
  26 & 0.648 & 0.806 & 0.600 & 0.815 \\ 
  27 & 0.510 & 0.794 & 0.415 & 0.710 \\ 
  28 & 0.434 & 0.933 & 0.309 & 0.567 \\ 
  29 & 0.615 & 0.769 & 0.544 & 0.802 \\ 
  30 & 0.630 & 0.827 & 0.583 & 0.790 \\ 
   \hline
\end{tabular}
\caption{ARB(Mx)/ARB(M12) for Mx = M11a, M11b, M1a, M1b for Case 1.  }
\label{table:arb1} 
\end{table}

\begin{table}[h!t!b!p!]
\centering
\begin{tabular}{ccccc}
 & M11a/M12 & M11b/M12 & M1a/M12 & M1b/M12 \\ 
  \hline
1 & 0.673 & 0.838 & 0.670 & 0.898 \\ 
  2 & 0.529 & 0.857 & 0.466 & 0.851 \\ 
  3 & 0.428 & 1.448 & 0.226 & 0.934 \\ 
  4 & 0.265 & 2.319 & 0.058 & 1.083 \\ 
  5 & 0.481 & 0.917 & 0.327 & 1.075 \\ 
  6 & 0.410 & 1.044 & 0.282 & 0.699 \\ 
  7 & 0.818 & 0.913 & 0.832 & 0.971 \\ 
  8 & 0.667 & 0.856 & 0.601 & 0.901 \\ 
  9 & 0.432 & 1.357 & 0.293 & 1.110 \\ 
  10 & 0.435 & 2.093 & 0.123 & 1.121 \\ 
  11 & 0.636 & 1.045 & 0.544 & 1.221 \\ 
  12 & 0.562 & 1.089 & 0.486 & 0.681 \\ 
  13 & 0.943 & 0.914 & 0.874 & 0.930 \\ 
  14 & 0.779 & 0.860 & 0.784 & 0.910 \\ 
  15 & 0.507 & 0.888 & 0.500 & 0.772 \\ 
  16 & 0.581 & 1.106 & 0.481 & 1.085 \\ 
  17 & 0.728 & 0.897 & 0.727 & 1.039 \\ 
  18 & 0.701 & 0.852 & 0.711 & 0.713 \\ 
  19 & 0.815 & 0.907 & 0.888 & 0.990 \\ 
  20 & 0.526 & 0.810 & 0.442 & 0.922 \\ 
  21 & 0.665 & 1.320 & 0.328 & 1.389 \\ 
  22 & 1.254 & 3.015 & 0.096 & 2.834 \\ 
  23 & 0.695 & 0.862 & 0.579 & 0.938 \\ 
  24 & 0.470 & 1.181 & 0.370 & 0.776 \\ 
  25 & 0.729 & 0.854 & 0.773 & 0.885 \\ 
  26 & 0.530 & 0.853 & 0.504 & 0.758 \\ 
  27 & 0.317 & 0.947 & 0.240 & 0.495 \\ 
  28 & 0.191 & 1.210 & 0.093 & 0.222 \\ 
  29 & 0.463 & 0.782 & 0.369 & 0.958 \\ 
  30 & 0.521 & 0.859 & 0.433 & 0.640 \\ 
   \hline
\end{tabular}
\caption{\label{table:asrb1}  ASRB(Mx)/ASRB(M12) for Mx = M11a, M11b, M1a, M1b for Case 1. }
\end{table}



\begin{table}[h!t!b!p!]
\centering
\begin{tabular}{ccccc}
& $p_1$ for  $\mu_i$  & $p_1$ for  $\theta_{ij}$ & $\tau_{21}$ for  $\mu_i$   & $\tau_{21}$ for  $\theta_{ij}$  \\
  \hline
1 & 0.1 & 0.1 & 0.1 & 0.05 \\ 
  2 &  &  &  & 0.1 \\ 
  3 &  &  &  & 0.2 \\ 
  4 &  &  & 0.2 & 0.05 \\ 
  5 &  &  &  & 0.1 \\ 
  6 &  &  &  & 0.2 \\ 
  7 &  & 0.2 & 0.1 & 0.05 \\ 
  8 &  &  &  & 0.1 \\ 
  9 &  &  &  & 0.2 \\ 
  10 &  &  & 0.2& 0.05 \\ 
  11 &  &  &  & 0.1 \\ 
  12 &  &  &  & 0.2 \\ 
  13 &  & 0.4& 0.1 & 0.05 \\ 
  14 &  &  &  & 0.1 \\ 
  15 &  &  &  & 0.2 \\ 
  16 &  &  & 0.2 & 0.05 \\ 
  17 &  &  &  & 0.1 \\ 
  18 &  &  &  & 0.2 \\ 
  19 & 0.2& 0.1 & 0.1 & 0.05 \\ 
  20 &  &  &  & 0.1 \\ 
  21 &  &  &  & 0.2 \\ 
  22 &  &  & 0.2 & 0.05 \\ 
  23 &  &  &  & 0.1 \\ 
  24 &  &  &  & 0.2 \\ 
  25 &  & 0.2 & 0.1 & 0.05 \\ 
  26 &  &  &  & 0.1 \\ 
  27 &  &  &  & 0.2 \\ 
  28 &  &  & 0.2 & 0.05 \\ 
  29 &  &  &  & 0.1\\ 
  30 &  &  &  & 0.2 \\ 
  31 &  & 0.4 & 0.1 & 0.05 \\ 
  32 &  &  &  & 0.1\\ 
  33 &  &  &  & 0.2 \\ 
  34 &  &  & 0.2& 0.05 \\ 
  35 &  &  &  & 0.1  \\ 
  36 &  &  &  & 0.2 \\ 
   \hline
\end{tabular}
\caption{\label{table:specs2}  Simulation specifications for Case 2.}
\end{table}

\begin{table}[h!t!b!p!]
\centering
\begin{tabular}{ccccc}
 & M11a/M12 & M11b/M12 & M1a/M12 & M1b/M12 \\ 
  \hline
1 & 0.935 & 0.967 & 0.942 & 0.995 \\ 
  2 & 0.749 & 0.847 & 0.725 & 0.860 \\ 
  3 & 0.559 & 0.724 & 0.502 & 0.750 \\ 
  4 & 0.888 & 0.937 & 0.894 & 0.962 \\ 
  5 & 0.741 & 0.835 & 0.723 & 0.831 \\ 
  6 & 0.617 & 0.773 & 0.551 & 0.812 \\ 
  7 & 0.937 & 0.972 & 0.936 & 0.983 \\ 
  8 & 0.726 & 0.853 & 0.709 & 0.846 \\ 
  9 & 0.541 & 0.710 & 0.487 & 0.704 \\ 
  10 & 0.874 & 0.929 & 0.887 & 0.942 \\ 
  11 & 0.755 & 0.865 & 0.739 & 0.841 \\ 
  12 & 0.563 & 0.752 & 0.498 & 0.740 \\ 
  13 & 0.932 & 0.957 & 0.951 & 0.966 \\ 
  14 & 0.768 & 0.891 & 0.742 & 0.880 \\ 
  15 & 0.593 & 0.819 & 0.525 & 0.736 \\ 
  16 & 0.906 & 0.940 & 0.923 & 0.947 \\ 
  17 & 0.804 & 0.918 & 0.754 & 0.865 \\ 
  18 & 0.629 & 0.846 & 0.556 & 0.755 \\ 
  19 & 0.951 & 0.979 & 0.959 & 0.993 \\ 
  20 & 0.731 & 0.831 & 0.713 & 0.843 \\ 
  21 & 0.531 & 0.697 & 0.483 & 0.720 \\ 
  22 & 0.898 & 0.928 & 0.924 & 0.955 \\ 
  23 & 0.741 & 0.833 & 0.702 & 0.847 \\ 
  24 & 0.572 & 0.750 & 0.519 & 0.779 \\ 
  25 & 0.932 & 0.964 & 0.943 & 0.974 \\ 
  26 & 0.764 & 0.911 & 0.764 & 0.880 \\ 
  27 & 0.535 & 0.714 & 0.465 & 0.658 \\ 
  28 & 0.893 & 0.946 & 0.886 & 0.935 \\ 
  29 & 0.805 & 0.877 & 0.749 & 0.835 \\ 
  30 & 0.544 & 0.760 & 0.492 & 0.701 \\ 
  31 & 0.936 & 0.981 & 0.917 & 0.962 \\ 
  32 & 0.788 & 0.894 & 0.775 & 0.856 \\ 
  33 & 0.616 & 0.819 & 0.521 & 0.740 \\ 
  34 & 0.925 & 0.960 & 0.919 & 0.965 \\ 
  35 & 0.787 & 0.935 & 0.727 & 0.886 \\ 
  36 & 0.663 & 0.871 & 0.572 & 0.773 \\ 
   \hline
\end{tabular}
\caption{\label{table:arb2}  ARB(Mx)/ARB(M12) for Mx = M11a, M11b, M1a, M1b for Case 2.}  
\end{table}


\begin{table}[h!t!b!p!]
\centering
\begin{tabular}{ccccc}
 & M11a/M12 & M11b/M12 & M1a/M12 & M1b/M12 \\ 
  \hline
1 & 0.861 & 0.907 & 0.892 & 0.929 \\ 
  2 & 0.554 & 0.764 & 0.520 & 0.671 \\ 
  3 & 0.366 & 0.621 & 0.255 & 0.518 \\ 
  4 & 0.814 & 0.925 & 0.816 & 0.905 \\ 
  5 & 0.723 & 0.803 & 0.707 & 0.752 \\ 
  6 & 0.486 & 0.626 & 0.375 & 0.648 \\ 
  7 & 0.888 & 1.016 & 0.887 & 0.981 \\ 
  8 & 0.509 & 0.737 & 0.473 & 0.619 \\ 
  9 & 0.250 & 0.444 & 0.166 & 0.345 \\ 
  10 & 0.840 & 0.886 & 0.823 & 0.853 \\ 
  11 & 0.619 & 0.747 & 0.569 & 0.646 \\ 
  12 & 0.292 & 0.490 & 0.217 & 0.516 \\ 
  13 & 0.858 & 0.964 & 0.873 & 0.902 \\ 
  14 & 0.566 & 0.836 & 0.481 & 0.671 \\ 
  15 & 0.332 & 0.695 & 0.259 & 0.456 \\ 
  16 & 0.883 & 0.915 & 0.862 & 0.931 \\ 
  17 & 0.667 & 0.874 & 0.617 & 0.697 \\ 
  18 & 0.419 & 0.702 & 0.330 & 0.520 \\ 
  19 & 0.871 & 0.946 & 0.940 & 0.970 \\ 
  20 & 0.595 & 0.771 & 0.579 & 0.679 \\ 
  21 & 0.380 & 0.615 & 0.252 & 0.548 \\ 
  22 & 0.851 & 0.933 & 0.826 & 0.904 \\ 
  23 & 0.624 & 0.777 & 0.570 & 0.694 \\ 
  24 & 0.450 & 0.652 & 0.405 & 0.680 \\ 
  25 & 0.892 & 0.980 & 0.892 & 0.961 \\ 
  26 & 0.560 & 0.712 & 0.616 & 0.710 \\ 
  27 & 0.277 & 0.443 & 0.203 & 0.363 \\ 
  28 & 0.759 & 0.926 & 0.723 & 0.809 \\ 
  29 & 0.506 & 0.635 & 0.515 & 0.633 \\ 
  30 & 0.292 & 0.499 & 0.245 & 0.485 \\ 
  31 & 0.823 & 0.994 & 0.725 & 0.907 \\ 
  32 & 0.626 & 0.864 & 0.603 & 0.691 \\ 
  33 & 0.355 & 0.747 & 0.260 & 0.491 \\ 
  34 & 0.875 & 0.925 & 0.793 & 0.884 \\ 
  35 & 0.725 & 0.937 & 0.668 & 0.803 \\ 
  36 & 0.471 & 0.824 & 0.411 & 0.615 \\ 
   \hline
\end{tabular}
\caption{\label{table:asrb2}  ASRB(Mx)/ASRB(M12) for Mx = M11a, M11b, M1a, M1b for Case 2. } 
\end{table}

 
\begin{table}[h!t!b!p!]
\centering
\begin{tabular}{ccccc}
& $p_1$  for $\mu_i$ & $p_2$ for $\theta_{ij}$ & $\tau_{21}$ for $\mu_i$ & $\tau_{11}$  for $\theta_{ij}$ \\
  \hline
  1 & 0.1 & 0.1 & 0.1 & 0.1 \\ 
  2 &  &  & 0.2 & 0.2 \\ 
  3 &  &  & 0.4 & 0.4 \\ 
  4 &  &  & 0.2 & 0.4 \\ 
  5 &  &  & 0.4 & 0.2 \\ 
  6 & 0.2 & 0.2 & 0.1 & 0.1 \\ 
  7 &  &  & 0.2 & 0.2 \\ 
  8 &  &  & 0.4 & 0.4 \\ 
  9 &  &  & 0.2 & 0.4 \\ 
  10 &  &  & 0.4 & 0.2 \\ 
  11 & 0.1 & 0.2 & 0.1 & 0.1 \\ 
  12 &  &  & 0.2 & 0.2 \\ 
  13 &  &  & 0.4 & 0.4 \\ 
  14 &  &  & 0.2 & 0.4 \\ 
  15 &  &  & 0.4 & 0.2 \\ 
  16 & 0.2 & 0.1 & 0.1 & 0.1 \\ 
  17 &  &  & 0.2 & 0.2 \\ 
  18 &  &  & 0.4 & 0.4 \\ 
  19 &  &  & 0.2 & 0.4 \\ 
  20 &  &  & 0.4 & 0.2 \\ 
   \hline
\end{tabular}
\caption{\label{table:specs3}  Simulation specifications  for  Case 3.}
\end{table}


\begin{table}[h!t!b!p!]
\centering
\begin{tabular}{ccccc}
Case & M11a/M12 & M11b/M12 & M1a/M12 & M1b/M12 \\ 
  \hline
1 & 1.031 & 1.003 & 1.050 & 0.993 \\ 
  2 & 0.849 & 0.923 & 0.844 & 0.880 \\ 
  3 & 0.696 & 0.821 & 0.655 & 0.762 \\ 
  4 & 0.690 & 0.817 & 0.680 & 0.896 \\ 
  5 & 0.864 & 0.937 & 0.828 & 0.855 \\ 
  6 & 0.972 & 0.964 & 1.001 & 0.969 \\ 
  7 & 0.784 & 0.890 & 0.766 & 0.848 \\ 
  8 & 0.680 & 0.838 & 0.636 & 0.819 \\ 
  9 & 0.634 & 0.815 & 0.619 & 0.908 \\ 
  10 & 0.827 & 0.947 & 0.783 & 0.874 \\ 
  11 & 1.017 & 0.982 & 1.025 & 0.978 \\ 
  12 & 0.820 & 0.906 & 0.786 & 0.848 \\ 
  13 & 0.677 & 0.840 & 0.619 & 0.841 \\ 
  14 & 0.697 & 0.783 & 0.671 & 0.937 \\ 
  15 & 0.822 & 0.927 & 0.778 & 0.836 \\ 
  16 & 0.988 & 0.983 & 1.006 & 0.965 \\ 
  17 & 0.867 & 0.925 & 0.852 & 0.866 \\ 
  18 & 0.670 & 0.824 & 0.609 & 0.749 \\ 
  19 & 0.615 & 0.754 & 0.596 & 0.794 \\ 
  20 & 0.913 & 0.971 & 0.904 & 0.929 \\ 
   \hline
\end{tabular}
\caption{\label{table:arb3}  ARB(Mx)/ARB(M12) for Mx = M11a, M11b, M1a, M1b for  Case 3.} 
\end{table}


\begin{table}[h!t!b!p!]
\centering
\begin{tabular}{ccccc}
 & M11a/M12 & M11b/M12 & M1a/M12 & M1b/M12 \\ 
  \hline
1 & 1.034 & 0.992 & 1.102 & 0.988 \\ 
  2 & 0.438 & 0.699 & 0.452 & 0.565 \\ 
  3 & 0.323 & 0.572 & 0.305 & 0.515 \\ 
  4 & 0.232 & 0.462 & 0.214 & 0.752 \\ 
  5 & 0.607 & 0.864 & 0.562 & 0.646 \\ 
  6 & 0.860 & 0.927 & 0.960 & 0.946 \\ 
  7 & 0.454 & 0.649 & 0.470 & 0.661 \\ 
  8 & 0.370 & 0.717 & 0.427 & 0.890 \\ 
  9 & 0.289 & 0.690 & 0.275 & 1.007 \\ 
  10 & 0.705 & 0.891 & 0.763 & 0.825 \\ 
  11 & 0.948 & 0.907 & 0.997 & 0.942 \\ 
  12 & 0.539 & 0.722 & 0.527 & 0.629 \\ 
  13 & 0.376 & 0.817 & 0.353 & 0.810 \\ 
  14 & 0.425 & 0.684 & 0.429 & 1.112 \\ 
  15 & 0.648 & 0.928 & 0.612 & 0.739 \\ 
  16 & 0.940 & 0.944 & 1.005 & 0.926 \\ 
  17 & 0.530 & 0.786 & 0.563 & 0.567 \\ 
  18 & 0.311 & 0.518 & 0.301 & 0.499 \\ 
  19 & 0.250 & 0.554 & 0.296 & 0.774 \\ 
  20 & 0.797 & 0.906 & 0.893 & 0.820 \\ 
   \hline
\end{tabular}
\caption{\label{table:asrb3}  ASRB(Mx)/ASRB(M12) for Mx = M11a, M11b, M1a, M1b for Case 3.} 
\end{table}





\begin{table}[h!t!b!p!]
\centering
\begin{tabular}{ccccc}
& $p_2$  for $\mu_i$  & $\tau_{11}$ for $\mu_i$  & $p_1$   for $\theta_{ij}$  &  $\tau_{21}$ for $\theta_{ij}$  \\
\hline
1 & 0.1 & 0.05 & 0.1 & 0.1 \\ 
  2 &  &  &  & 0.2 \\ 
  3 &  &  &  & 0.4 \\ 
  4 &  &  & 0.2 & 0.1 \\ 
  5 &  &  &  & 0.2 \\ 
  6 &  &  &  & 0.4 \\ 
  7 &  & 0.1 & 0.1 & 0.1 \\ 
  8 &  &  &  & 0.2 \\ 
  9 &  &  &  & 0.4 \\ 
  10 &  &  & 0.2 & 0.1 \\ 
  11 &  &  &  & 0.2 \\ 
  12 &  &  &  & 0.4 \\ 
  13 &  & 0.2 & 0.1 & 0.1 \\ 
  14 &  &  &  & 0.2 \\ 
  15 &  &  &  & 0.4 \\ 
  16 &  &  & 0.2 & 0.1 \\ 
  17 &  &  &  & 0.2 \\ 
  18 &  &  &  & 0.4 \\ 
  19 & 0.2 & 0.05 & 0.1 & 0.1 \\ 
  20 &  &  &  & 0.2 \\ 
  21 &  &  &  & 0.4 \\ 
  22 &  &  & 0.2 & 0.1 \\ 
  23 &  &  &  & 0.2 \\ 
  24 &  &  &  & 0.4 \\ 
  25 &  & 0.1 & 0.1 & 0.1 \\ 
  26 &  &  &  & 0.2 \\ 
  27 &  &  &  & 0.4 \\ 
  28 &  &  & 0.2 & 0.1 \\ 
  29 &  &  &  & 0.2 \\ 
  30 &  &  &  & 0.4 \\ 
  31 &  & 0.2 & 0.1 & 0.1 \\ 
  32 &  &  &  & 0.2 \\ 
  33 &  &  &  & 0.4 \\ 
  34 &  &  & 0.2 & 0.1 \\ 
  35 &  &  &  & 0.2 \\ 
  36 &  &  &  & 0.4 \\ 
   \hline
\end{tabular}
\caption{\label{table:specs4}  Simulation specifications   for Case 4.}
\end{table}
\pagebreak

\begin{table}[h!t!b!p!]
\centering
\begin{tabular}{ccccc}
 & M11a/M12 & M11b/M12 & M1a/M12 & M1b/M12 \\ 
  \hline
1 & 1.747 & 1.321 & 1.353 & 1.095 \\ 
  2 & 1.468 & 1.054 & 1.175 & 0.805 \\ 
  3 & 1.032 & 0.684 & 0.815 & 0.569 \\ 
  4 & 1.567 & 1.207 & 1.270 & 1.054 \\ 
  5 & 1.456 & 1.023 & 1.122 & 0.815 \\ 
  6 & 0.923 & 0.587 & 0.775 & 0.489 \\ 
  7 & 1.264 & 1.109 & 1.059 & 1.006 \\ 
  8 & 1.060 & 0.931 & 0.865 & 0.859 \\ 
  9 & 0.904 & 0.762 & 0.749 & 0.655 \\ 
  10 & 1.190 & 1.095 & 1.025 & 0.964 \\ 
  11 & 1.006 & 0.861 & 0.804 & 0.743 \\ 
  12 & 0.882 & 0.701 & 0.760 & 0.618 \\ 
  13 & 0.808 & 0.950 & 0.687 & 0.815 \\ 
  14 & 0.770 & 0.924 & 0.651 & 0.823 \\ 
  15 & 0.675 & 0.793 & 0.548 & 0.779 \\ 
  16 & 0.818 & 0.951 & 0.673 & 0.793 \\ 
  17 & 0.824 & 0.921 & 0.644 & 0.857 \\ 
  18 & 0.689 & 0.766 & 0.582 & 0.789 \\ 
  19 & 1.365 & 1.181 & 1.183 & 1.044 \\ 
  20 & 1.265 & 1.055 & 1.088 & 0.921 \\ 
  21 & 1.023 & 0.805 & 0.875 & 0.667 \\ 
  22 & 1.450 & 1.191 & 1.216 & 1.081 \\ 
  23 & 1.212 & 0.928 & 1.009 & 0.850 \\ 
  24 & 0.853 & 0.654 & 0.747 & 0.607 \\ 
  25 & 0.942 & 0.971 & 0.832 & 0.895 \\ 
  26 & 0.903 & 0.957 & 0.797 & 0.862 \\ 
  27 & 0.844 & 0.854 & 0.710 & 0.838 \\ 
  28 & 1.005 & 0.995 & 0.903 & 0.941 \\ 
  29 & 0.938 & 0.920 & 0.821 & 0.863 \\ 
  30 & 0.848 & 0.781 & 0.733 & 0.705 \\ 
  31 & 0.805 & 1.056 & 0.695 & 0.823 \\ 
  32 & 0.731 & 0.914 & 0.645 & 0.763 \\ 
  33 & 0.570 & 0.846 & 0.502 & 0.875 \\ 
  34 & 0.803 & 1.031 & 0.679 & 0.833 \\ 
  35 & 0.748 & 0.975 & 0.616 & 0.873 \\ 
  36 & 0.683 & 0.845 & 0.590 & 0.901 \\ 
   \hline
\end{tabular}
\caption{\label{table:arb4}  ARB(Mx)/ARB(M12) for Mx = M11a, M11b, M1a, M1b   for Case 4.} 
\end{table}

\pagebreak

\begin{table}[h!t!b!p!]
\centering
\begin{tabular}{ccccc}
 & M11a/M12 & M11b/M12 & M1a/M12 & M1b/M12 \\ 
  \hline
1 & 2.032 & 1.319 & 1.545 & 1.097 \\ 
  2 & 1.739 & 0.960 & 1.344 & 0.753 \\ 
  3 & 1.137 & 0.663 & 0.777 & 0.537 \\ 
  4 & 1.827 & 1.212 & 1.402 & 0.998 \\ 
  5 & 1.682 & 0.949 & 1.254 & 0.710 \\ 
  6 & 0.940 & 0.389 & 0.807 & 0.318 \\ 
  7 & 0.874 & 0.911 & 0.725 & 0.842 \\ 
  8 & 0.675 & 0.877 & 0.584 & 0.957 \\ 
  9 & 0.660 & 0.670 & 0.504 & 0.762 \\ 
  10 & 0.799 & 0.893 & 0.693 & 0.920 \\ 
  11 & 0.778 & 0.797 & 0.684 & 0.808 \\ 
  12 & 0.747 & 0.718 & 0.817 & 0.777 \\ 
  13 & 0.611 & 1.583 & 0.444 & 0.628 \\ 
  14 & 0.370 & 1.471 & 0.298 & 1.143 \\ 
  15 & 0.325 & 1.145 & 0.223 & 1.299 \\ 
  16 & 0.588 & 1.846 & 0.429 & 0.725 \\ 
  17 & 0.399 & 1.258 & 0.294 & 1.060 \\ 
  18 & 0.568 & 1.114 & 0.424 & 1.324 \\ 
  19 & 1.363 & 1.063 & 1.288 & 1.005 \\ 
  20 & 1.226 & 0.948 & 1.002 & 0.887 \\ 
  21 & 0.860 & 0.670 & 0.679 & 0.635 \\ 
  22 & 1.330 & 1.010 & 1.076 & 0.947 \\ 
  23 & 1.216 & 0.912 & 1.007 & 0.828 \\ 
  24 & 0.859 & 0.590 & 0.822 & 0.574 \\ 
  25 & 0.631 & 0.798 & 0.586 & 0.743 \\ 
  26 & 0.531 & 0.874 & 0.472 & 0.842 \\ 
  27 & 0.468 & 0.873 & 0.426 & 1.120 \\ 
  28 & 0.676 & 0.777 & 0.637 & 0.740 \\ 
  29 & 0.593 & 0.817 & 0.594 & 0.900 \\ 
  30 & 0.606 & 0.758 & 0.610 & 0.882 \\ 
  31 & 0.600 & 1.557 & 0.559 & 0.655 \\ 
  32 & 0.399 & 0.849 & 0.385 & 0.628 \\ 
  33 & 0.205 & 0.950 & 0.160 & 1.059 \\ 
  34 & 0.557 & 1.944 & 0.450 & 0.679 \\ 
  35 & 0.495 & 1.245 & 0.429 & 1.067 \\ 
  36 & 0.451 & 1.073 & 0.374 & 1.538 \\ 
   \hline
\end{tabular}
\caption{\label{table:asrb4}  ASRB(Mx)/ASRB(M12) for Mx = M11a, M11b, M1a, M1b  for Case 4.} 
\end{table}

\clearpage

\begin{table}[h!t!b!p!]
\centering
\begin{tabular}{rrrr}
  & M12/M1a & MSA/M1a & MBR/M1a \\ 
   \hline
1 & 1.036 & 1.104 & 2.271 \\ 
  2 & 1.252 & 1.354 & 2.145 \\ 
  3 & 2.030 & 2.127 & 2.025 \\ 
  4 & 1.064 & 1.224 & 1.838 \\ 
  5 & 1.328 & 1.570 & 1.601 \\ 
  6 & 1.068 & 1.363 & 1.450 \\ 
  7 & 1.067 & 1.401 & 1.441 \\ 
  8 & 1.300 & 1.946 & 1.173 \\ 
  9 & 1.703 & 3.423 & 1.111 \\ 
  10 & 1.025 & 1.080 & 2.342 \\ 
  11 & 1.000 & 1.345 & 2.070 \\ 
  12 & 1.000 & 1.980 & 2.025 \\ 
  13 & 1.072 & 1.188 & 1.931 \\ 
  14 & 1.322 & 1.631 & 1.724 \\ 
  15 & 1.866 & 2.623 & 1.534 \\ 
  16 & 1.068 & 1.349 & 1.497 \\ 
  17 & 1.286 & 2.009 & 1.223 \\ 
  18 & 1.651 & 3.435 & 1.088 \\ 
   \hline
\end{tabular}
\caption{\label{table:arb6}  ARB(Mx)/ARB(M1a) for Mx = M12, MSA, MBR for Case 5.}
\end{table}

\begin{table}[h!t!b!p!]
\centering
\begin{tabular}{rrrr}
 & M12/M1a & MSA/M1a & MBR/M1a \\ 
  \hline
1 & 1.137 & 1.495 & 4.361 \\ 
  2 & 1.853 & 3.410 & 3.433 \\ 
  3 & 3.899 & 14.065 & 2.946 \\ 
  4 & 1.133 & 1.941 & 2.535 \\ 
  5 & 1.846 & 4.889 & 1.981 \\ 
  6 & 1.075 & 2.338 & 1.754 \\ 
  7 & 1.062 & 2.342 & 1.698 \\ 
  8 & 1.488 & 5.172 & 1.111 \\ 
  9 & 2.408 & 19.540 & 0.962 \\ 
  10 & 1.187 & 1.468 & 4.465 \\ 
  11 & 1.000 & 3.848 & 3.364 \\ 
  12 & 1.000 & 13.375 & 3.168 \\ 
  13 & 1.208 & 1.876 & 2.907 \\ 
  14 & 1.734 & 4.924 & 2.160 \\ 
  15 & 3.239 & 17.360 & 1.831 \\ 
  16 & 1.135 & 2.197 & 1.898 \\ 
  17 & 1.514 & 5.851 & 1.232 \\ 
  18 & 2.339 & 20.498 & 0.983 \\ 
   \hline
\end{tabular}
\caption{\label{table:asrb6}  ASRB(Mx)/ASRB(M1a) for Mx = M12, MSA, MBR for  Case 5.}
\end{table}


\begin{table}[h!t!b!p!]
\centering
\begin{tabular}{rrrr}
 & M12/M1a & MSA/M1a & MBR/M1a \\ 
  \hline
 1 & 0.967 & 1.105 & 2.142 \\ 
  2 & 0.960 & 1.210 & 1.481 \\ 
  3 & 1.004 & 1.773 & 0.927 \\ 
  4 & 1.112 & 2.849 & 0.777 \\ 
  5 & 1.102 & 4.578 & 0.620 \\ 
  6 & 1.170 & 8.148 & 0.546 \\ 
  7 & 0.954 & 1.065 & 2.168 \\ 
  8 & 0.967 & 1.200 & 1.537 \\ 
  9 & 0.986 & 1.785 & 0.981 \\ 
  10 & 1.097 & 2.802 & 0.780 \\ 
  11 & 1.000 & 4.649 & 0.638 \\ 
  12 & 1.000 & 8.230 & 0.566 \\ 
   \hline
\end{tabular}
\caption{\label{table:arb5} ARB(Mx)/ARB(M1a) for Mx = M12, MSA, MBR for Case 6.}
\end{table}

\begin{table}[h!t!b!p!]
\centering
\begin{tabular}{rrrr}
 & M12/M1a & MSA/M1a & MBR/M1a \\ 
  \hline
1 & 0.912 & 1.169 & 4.610 \\ 
  2 & 0.918 & 1.459 & 2.426 \\ 
  3 & 0.950 & 2.980 & 0.863 \\ 
  4 & 1.175 & 7.155 & 0.583 \\ 
  5 & 1.202 & 19.205 & 0.374 \\ 
  6 & 1.289 & 60.459 & 0.293 \\ 
  7 & 0.908 & 1.083 & 4.745 \\ 
  8 & 0.882 & 1.431 & 2.418 \\ 
  9 & 0.905 & 2.931 & 0.954 \\ 
  10 & 1.217 & 7.528 & 0.599 \\ 
  11 & 1.000 & 18.028 & 0.382 \\ 
  12 & 1.000 & 59.006 & 0.311 \\ 
   \hline
\end{tabular}
\caption{\label{table:asrb5}  ASRB(Mx)/ASRB(M1a) for Mx = M12, MSA, MBR for Case 6.}
\end{table}

\begin{table}[htbp]
\centering
\begin{tabular}{c|c|cc|cc}
   & &   \multicolumn{2}{c}{ARB}  &   \multicolumn{2}{|c}{ASRB}  \\ 
  Case & total \#subcases & \#\{M11a$>$M11b\}&    \#\{M1a$>$M1b\} &  \#\{M11a$>$M11b\}&    \#\{M1a$>$M1b\}  \\ 
\hline
   1 & 30 & 28 & 30&  29&  30 \\ 
   2 & 36 & 36 & 36&  36&  36 \\ 
  3 & 20 &16 & 16 & 18&  15 \\ 
  4 & 36 & 15 & 17 & 23&  23 \\ 
\hline
Total & 122 & 95 & 99 &106 & 104 \\
\end{tabular}
\caption{\label{table:counts2} For each case:  \#\{Mx$>$My\}  is the number of times  A(Mx)/A(M12) is smaller than  A(My)/A(M12), A=ARB, ASRB; \{x,y=1a,1b, 11a, 11b\}. }
\end{table}


\begin{thebibliography}{9}
\bibitem[Malec and Sedransk(1992)]{mas92} Malec D and Sedransk J.  (1992). Bayesian methodology for combining the results from different experiments when the specifications for pooling are uncertain.  \emph{Biometrika}, 79(3), 593 - 601.
\end{thebibliography}
\end{document}